

\input harvmac

\def \tt {{\tilde \t}}
\def \G {\Gamma} \def \k1 {{1\over
k}} \def \bh { {\bar h} } \def \ov { \over }
  \def \B { { \bar B }}

\def \O {\Omega }

\def \ra {\rightarrow}
\def \W { {\bar \o }}
 \def \in {\int d^2 z  }
\def \half{{\textstyle{1\over 2}}}

\def \D {\Delta}

\def \a {\alpha}
\def \b {\beta}
\def \hi {\chi}
\def \ps {\psi }
\def \sh {{\rm sinh \ }}
\def \Tr {{\ \rm Tr \ }}

\def \ln {{\rm \ ln \  }}
\def \det {{\ \rm det \ }}

\def \th {{\rm tanh  }}
\def \l {\lambda}
\def \1p {{1\over  \pi }}
\def \2p {{{1\over  2\pi }}}
\def \4p {{ {1\over 4 \pi }}}
\def \8p {{{1\over 8 \pi }}}
\def \P^* { P^{\dag } }
\def \p {\phi}
\def \M {{\cal M}}
\def \m {\mu }
\def \n {\nu}
\def \ep {\epsilon}
\def\g {\gamma}
\def \r {\rho}
\def \k {\kappa }
\def \d {\delta}
\def \o {\omega}
\def \s {\sigma}
\def \t {\theta}

\def \Gmn {G_{\mu \nu}}
\def \fourth {{\textstyle{1\over 4}}}
\def \third {{\textstyle{1\over 3}}}

\def \e#1 {{{\rm e}^{#1}}}
\def \const {{\rm const }}

\def \eq#1 {\eqno {(#1)}}
\def \sm {sigma model\ }\def \B  {{ \tilde B }}

\def \dm {\del_\m}
\def \dn {\del_\n}
\def \go {G^{(0)}}
\def \cab {{\cal C}^{AB}}
\def \bd  {{ \bar \del }}
\def \sl {SL(2,R)/U(1)}
\def \D  {\Delta }
\def \M {{\cal M}}
\def \Gwzw  {gauged WZW theory \ }
\def \tj   {{\tilde {\cal J}}}
\def \V { {\tilde V } }
\def \E {{ \tilde E}}
\def \bH {{\bar H}}
\def \tH {{ \tilde H}}
\def \cj   {{\cal J}}

\def \B  {{ \tilde B }}

\def \dm {\del_\m}
\def \dn {\del_\n}
\def \go {G^{(0)}}

\def \bd  { \bar \del }
\def \D  {\Delta }
\def \tl {{ \theta }_L}
\def \tR {{ \theta }_R}
\def \ov {\over }
\def \kg { k + \ha c_G}
\def \kh { k + \ha c_H}
\def \A  { {\bar A} }
\def \tth {\tilde h}
\def \SWZW {supersymmetric WZW theory \ }
\def \C {{\cal C }}

\def \o {\omega}

\def \p {\phi}
\def \ep {\epsilon}
\def \s {\sigma}

\def \gr {\rho}
\def \r {\rho}
\def \d {\delta}
\def \l {\lambda}
\def \m {\mu}
\def \g {\gamma}
\def \n {\nu}

\def \Gmn {G_{\mu \nu}}
\def \fourth {{1\over 4}}
\def \third {{1\over 3}}
\def \e#1 {{{\rm e}^{#1}}}
\def \const {{\rm const }}

\def \vp {\varphi}

\def \hg {{\hat g}}
\def \B {{\bar B}}
\def \H {{\cal H}}

\def \tg {{\tilde g}}
\def \half { { 1\over 2 }}
\def  \ee {{\rm e}^}
\def \J {\bar J }
\def\np {  Nucl. Phys. }
\def \pl { Phys. Lett. }
\def \mpl { Mod. Phys. Lett. }
\def \prl { Phys. Rev. Lett. }
\def \pr  { Phys. Rev. }
\def \ap  { Ann. Phys. }
\def \cmp { Commun. Math. Phys. }
\def \ijmp { Int. J. Mod. Phys. }
\Title{\vbox
{\baselineskip14pt{\hbox{CERN-TH.6804/93}}{\hbox{hep-th/9302083}}}}
{\vbox{\centerline {Conformal  Sigma  Models }\vskip2pt
 \centerline{Corresponding  to Gauged  } \vskip2pt
\centerline{ Wess-Zumino-Witten Theories}
}}
\centerline{ A.A. Tseytlin\footnote{$^*$}{
On leave from Blackett Laboratory, Imperial College, London SW7 2BZ, U.K.
and P.N. Lebedev Physics
Institute, Moscow, Russia.  $ \ \ $
E-mail: tseytlin@surya3.cern.ch and tseytlin@ic.ac.uk} }
\centerline {\it Theory Division, CERN}
\centerline {\it
CH-1211 Geneva 23, Switzerland}

\bigskip

\centerline {\bf Abstract}
\baselineskip12pt
\noindent
We develop a field-theoretical approach to  determination  of the background
target space fields
corresponding to  general $G/H$ coset conformal theories described by  gauged
WZW models.
The basic idea is to identify the effective action  of a gauged WZW theory
with the effective
action of  a sigma model.  The derivation of the quantum  effective action in
the  gauged WZW
theory is presented in detail,  both in the bosonic and in the supersymmetric
cases. We explain why
and how one can truncate the effective action by omitting  most of the
non-local terms (thus
providing  a justification for some  previous suggestions). The resulting
metric, dilaton and the
antisymmetric tensor are non-trivial functions of $1/k$ (or $\alpha'$) and
represent a large class
of conformal sigma models. The exact expressions for the fields in the
sypersymmetric case are equal
to  the leading order (`semiclassical') bosonic expressions (with no shift of
$k$). An  explicit
form in which we  find the  sigma model  couplings  makes it possible to prove
that the metric and
the  dilaton are equivalent to the fields which  are obtained  in the operator
approach, i.e. by
identifying  the $L_0$-operator of the conformal theory with a Klein-Gordon
operator in a
background. The metric can be considered as a `deformation'  of an invariant
metric  on the coset
space $G/H$ and the dilaton  can be in general represented  in terms  of  the
logarithm of the ratio
of the determinants of the `deformed' and `round' metrics.

\bigskip
\noindent
{CERN-TH.6804/93}

\Date
 {February 1993}

\noblackbox
\baselineskip 20pt plus 2pt minus 2pt

\lref \bep {C. Becchi and O. Piguet, \np B315(1989)153. }
\lref \nov {S.P. Novikov,  Sov. Math. Dokl. 37(1982)3. }
\lref \fuch { J. Fuchs, \np B286(1987)455; \np B318(1989)631. }
\lref \sus { R. Rohm,  \pr D32(1985)2845.}
 \lref \suss { H.W. Braden, \pr D33(1986)2411.}
\lref   \red { A.N. Redlich and H.J. Schnitzer, \pl B167(1986)315;
B193(1987)536 (E);
E. Bergshoeff, S. Randjbar - Daemi, A. Salam, H. Sarmadi and E. Sezgin,
\np B269(1986)77;  A. Ceresole,
 A. Lerda, P. Pizzochecco
 and
P. van Nieuwenhuizen, \pl
 B189(1987)34.}
 \lref \div  { P. Di Vecchia,  V. Knizhnik, J. Peterson and P. Rossi, \np
B253(1985)701.}

\lref \schn {  H. Schnitzer, \np B324(1989)412.  }

\lref \all { R.W. Allen, I. Jack and D.R.T. Jones, Z. Phys. C41(1988)323. }

\lref \nem {D. Nemeschansky and S. Yankielowicz, \prl 54(1985)620; 54(1985)1736
(E).}

\lref \bep { C. Becchi and O. Piguet, \np B315(1989)153. }

\lref \ks {Y. Kazama and H. Suzuki, \np B321(1989)232; \pl B216(1989)112.}

\lref \mor {A.Yu. Morozov, A.M. Perelomov, A.A. Rosly, M.A. Shifman and A.V.
Turbiner, \ijmp
A5(1990)803.}

 \lref \tur {A.V. Turbiner, \cmp 118(1988)467;  M.A. Shifman and A.V. Turbiner,
\cmp 126(1989)347;
M.A. Shifman, \ijmp A4(1989)2897.}

\lref \hal { M.B. Halpern and E.B. Kiritsis, \mpl A4(1989)1373; A4(1989)1797
(E).}

\lref \haly {M.B. Halpern and   J.P. Yamron,  Nucl.Phys.B332(1990)411;
Nucl.Phys.
B351(1991)333.}
\lref \halp { M.B. Halpern, E.B. Kiritsis, N.A. Obers, M. Porrati and J.P.
Yamron,
\ijmp A5(1990)2275;
  A.Yu. Morozov,  M.A. Shifman and A.V. Turbiner, \ijmp
A5(1990)2953;
A. Giveon, M.B. Halpern, E.B. Kiritsis and  N.A. Obers,
\np B357(1991)655.}
\lref \bpz {A.A. Belavin, A.M. Polyakov and A.B. Zamolodchikov, \np
B241(1984)333. }
\lref \efr {     S. Elitzur, A. Forge and E. Rabinovici, \np B359 (1991)
581;
 G. Mandal, A. Sengupta and S. Wadia, Mod. Phys. Lett. A6(1991)1685. }
\lref \sak {K. Sakai, Kyoto preprint, KUNS-1141-1992. }
\lref \ver {H. Verlinde, \np B337(1990)652.}
\lref \gwz {      K. Bardakci, E. Rabinovici and
B. S\"aring, \np B299(1988)157;
 K. Gawedzki and A. Kupiainen, \pl B215(1988)119;
\np B320(1989)625. }

\lref \sen {A. Sen, preprint TIFR-TH-92-57. }

\lref \bcr {K. Bardakci, M. Crescimanno and E. Rabinovici, \np
B344(1990)344. }
\lref \Jack {I. Jack, D.R.T.  Jones and J. Panvel, Liverpool preprint
LTH-277 (1992). }
\lref \zam  { Al. B. Zamolodchikov, preprint ITEP 87-89. }

\lref \hor {J. Horne and G. Horowitz, \np B368(1992)444. }
\lref \tse { A.A. Tseytlin, \pl B264(1991)311. }
\lref \gwzw  { P. Di Vecchia and P. Rossi, \pl  B140(1984)344;
 P. Di Vecchia, B. Durhuus  and J. Petersen, \pl  B144(1984)245.}
\lref \oal { O. Alvarez, \np B238(1984)61. }

\lref \ishi { N. Ishibashi, M.  Li and A. Steif, \prl 67(1991)3336. }
\lref  \kumar  { M. Ro\v cek and E. Verlinde, \np B373(1992)630; A. Kumar,
preprint CERN-TH.6530/92;
 S. Hussan and A. Sen,  preprint  TIFR-TH-92-61;  D. Gershon,
preprint TAUP-2005-92; X. de la Ossa and F. Quevedo, preprint NEIP92-004; E.
Kiritsis, preprint LPTENS-92-29. }

\lref \rocver { A. Giveon and M. Ro\v cek, \np B380(1992)128. }
\lref \frts {E.S. Fradkin and A.A. Tseytlin, \np B261(1985)1. }
\lref \mplt {A.A. Tseytlin, \mpl A6(1991)1721. }
\lref\bn {I. Bars and D. Nemeschansky, \np B348(1991)89.}
\lref \shif { M.A. Shifman, \np B352(1991)87.}
\lref\wittt { E. Witten, \cmp 121(1989)351; G. Moore and N. Seiberg, \pl
B220(1989)422.} \lref \chernsim { E. Guadagnini, M. Martellini and M.
Mintchev, \np B330(1990)575;
L. Alvarez-Gaume, J. Labastida and A. Ramallo, \np B354(1990)103;
G. Giavarini, C.P. Martin and F. Ruiz Ruiz, \np B381(1992)222; preprint
LPTHE-92-42.}
\lref \shifley { H. Leutwyler and M.A. Shifman, \ijmp A7(1992)795. }
\lref \polwig { A.M. Polyakov and P.B. Wiegman, \pl B131(1984)121; \pl
B141(1984)223.  }
\lref \polles { A. Polyakov, in: {\it Fields, Strings and Critical Phenomena},
  Proc. of Les Houches 1988,  eds.  E. Brezin and J. Zinn-Justin
(North-Holland,1990).   }
\lref \kutas {
D. Kutasov, \pl B233(1989)369.} \lref \karabali { D. Karabali, Q-Han Park, H.J.
Schnitzer and
Z. Yang, \pl B216(1989)307;  D. Karabali and H.J. Schnitzer, \np B329(1990)649.
}
\lref \ginq {P. Ginsparg and F. Quevedo,  \np B385(1992)527. }
\lref \gko  {K. Bardakci and M.B. Halpern, \pr D3(1971)2493;
 M.B. Halpern, \pr D4(1971)2398;   P. Goddard,
A. Kent and D. Olive, \pl B152(1985)88; \cmp 103(1986)303;  V. Kac and I.
Todorov, \cmp
102(1985)337.  }
\lref \dvv  { R. Dijkgraaf, H. Verlinde and E. Verlinde, \np B371(1992)269. }
\lref \kniz {  V. Knizhnik and A. Zamolodchikov, \np B247(1984)83. }

\lref \witt { E. Witten, \cmp 92(1984)455.}
\lref \wit { E. Witten, \pr D44(1991)314.}
\lref \anton { I. Antoniadis, C. Bachas, J. Ellis and D.V. Nanopoulos, \pl B211
(1988)393.}
\lref \bsfet {I. Bars and  K. Sfetsos, \pr D46(1992)4510; preprint
USC-92/HEP-B3 (1992);  K. Sfetsos,  preprint USC-92/HEP-S1 (1992).}

\lref \ts  {A.A.Tseytlin, \pl B268(1991)175. }
\lref \kir {{E. Kiritsis, \mpl A6(1991)2871.} }

\lref \shts {A.S. Schwarz and A.A. Tseytlin,  preprint Imperial/TP/92-93/01
(1992). }
\lref \bush { T.H. Buscher, \pl B201(1988)466.  }

\lref \plwave { D. Amati and C. Klim\v cik, \pl B219(1989)443; G. Horowitz and
A. Steif, \prl 64(1990)260.}
\lref\bsft { I. Bars and K. Sfetsos, \mpl A7(1992)1091;
 \pr D46(1992)4495;
I. Bars, preprint USC-92/HEP-B5. }
\lref\bs { I. Bars and K. Sfetsos,  preprint USC-93/HEP-B1 (1993). }
\lref\bst { I. Bars,  K. Sfetsos and A.A. Tseytlin, unpublished. }
\lref\bb  { I. Bars, \np B334(1990)125. }
\lref \tsw  { A.A. Tseytlin, preprint Imperial/TP/92-93/10 (1992).}
\lref \ger { A. Gerasimov, A. Morozov, M. Olshanetsky, A. Marshakov and S.
Shatashvili, \ijmp
A5(1990)2495. }
\lref \sch { K. Schoutens, A. Sevrin and P. van Nieuwenhuizen,
in: Proc. of the Stony Brook Conference {\it `Strings and Symmetries 1991'},
p.558  (World
Scientific, Singapore, 1992).} \lref \boer { J. de Boer and J. Goeree, Utrecht
preprint THU-92/33. }
\lref \dev { C. Destri and H.J. De Vega, \pl B208(1988)255. }
\lref  \noj {S. Nojiri, \pl B271(1992)41. }
\lref \swz  { E. Witten, \np B371(1992)191;
T. Nakatsu, Progr. Theor. Phys. 87(1992)795. }
 \lref \bsf { I. Bars and K. Sfetsos, \pl B277(1992)269. }

\lref \br {A. Barut and R. Raczka, ``Theory of Group Representations and
Applications", p.120
 (PWN, Warszawa 1980). }
\lref \jjmo {I. Jack, D.R.T. Jones, N.Mohammedi and H. Osborn, \np
B332(1990)359;
C.M. Hull and B. Spence, \pl B232(1989)204. }

\lref \tttt { A.A. Tseytlin,  preprint Imperial/TP/92-93/7 (1992), \pr
D47(1993) no.8.}
\lref \per { V. Novikov, M. Shifman, A. Vainshtein and V. Zakharov, \pl
B139(1984)389;
A. Morozov, A. Perelomov and M. Shifman, \np B248(1984)279;
M.C. Prati and A.M. Perelomov, \np B258(1985)647. }
\lref \alv {
L. Alvarez-Gaum\'e,   D. Freedman and S. Mukhi, \ap 134(1981)85;
L. Alvarez-Gaum\'e, \np B184(1981)180. }
\lref \gr {  M.T. Grisaru, A. van de Ven and D. Zanon, \np B277(1986)409. }

\newsec {Introduction }

Given a conformal theory formulated in the operator approach \bpz\  it is not
{\it a priori }
clear which  is a $2d$ field theory (`sigma model')  which corresponds to it
(if such
 correspondence is possible at all).
To have  a sigma-model  interpretation  (at least in some limit) $is$ important
 in the case
when a conformal theory is used to represent a solution of string theory.
A large class of (super)conformal theories   based on
 the $G/H$ coset construction \gko\  can be   described in terms of  gauged
Wess-Zumino-Witten
theories \gwz\karabali. Both   compact \ks\ and non-compact \bb\bn\  coset
models
describe   most of the known Euclidean  (`internal' space) and  Minkowski
(black hole  and
cosmological) string  solutions.

The first example of a \sm interpretation of a  gauged WZW model was  given in
 \bcr\wit\
for the case of the $SU(2)/U(1)$ or $SL(2,R)/U(1)$  theory. It was found that
the \sm metric is
different \bcr\ from the
standard invariant metric on the coset space and that the \sm contains  also a
non-trivial  dilaton
coupling \wit. These two facts are, of course, not unrelated:
  the dilaton must be present in order to satisfy the conformal invariance
condition  given that
the invariant metric on a  homogeneous  space  has a non-trivial Ricci tensor
\wit.  The idea in
\bcr\wit\ was to
 eliminate the
$2d$ gauge field from the classical action of   gauged WZW model.  Since this
was done at the
semiclassical level only,  the resulting \sm was conformal only in the  leading
order (in $\a'$ or
$1/k$)  approximation. It was suggested in \dvv\ that the exact expressions for
the metric and the
dilaton can be  obtained  by using the `operator approach', i.e.  by
interpreting the $L_0$-operator
of the corresponding coset theory as a Klein-Gordon operator in a background.
The expressions
proposed in \dvv\ (which explicitly  depended on $\alpha'$) were, in fact,
found   to be solutions
of the \sm conformal invariance conditions  up to four loop orders \ts\Jack.

Since the operator approach  to  determination   of the  background fields is
rather  heuristic,
 being based on a  number of implicit assumptions (and   do not allowing  one
to
 compute  the antisymmetric tensor coupling in a straightforward way)    it is
desirable  to
have a  direct  field-theoretic  method of derivation of a \sm corresponding to
a  coset
theory  which generalizes the  idea of  \wit\ to all orders in $1/k$.   Such a
method was  recently
suggested in \tsw\ and considered also in \bs. The  main  point  is  first to
replace   the
classical  gauged WZW  action by the  exact  effective one   and {\it then  }
eliminate the gauge
field.    Since the \Gwzw  is `exactly soluble',  its  effective action can  be
found  explicitly
\tsw. The aim of  the present  paper is to  clarify and develop further this
approach. In
particular, we shall generalize the analysis of \tsw\ to the case of \SWZW and
derive the general
expressions for the  metric, dilaton and antisymmetric tensor for an arbitrary
$G/H$ model
(justifying and extending the results of \tsw\bs). The explicit form in which
we shall find
the \sm couplings will make it possible to  check that the resulting  metric
and dilaton are the same
that appear (in  more  abstract form) in the operator approach \dvv\bsfet\sak.
 We shall also  prove
in general  that the dilaton   can be   expressed    essentially   in terms
the logarithm of the
determinant of the metric (confirming previous   suggestions
 \kir\bsfet\bs).\foot {  The   fact    that  the product $\sqrt G \ {\rm
e}^{-2 \p}$  is  $k$-independent   was  observed  in the $SL(2,R)/U(1)$ case
in \dvv\kir. It  was   further checked on a number
of  non-trivial  $G/H$ models,  formulated  as a general statement
  and argued for using path integral measure considerations in \bsfet\bs.}

Let us   first    explain    the   basic idea  of our approach.   To give a
\sm
interpretation  to a gauged WZW theory one, should, in principle,  should fix
a gauge
and  integrate
out   the gauge field. In practice, this is rather difficult to do since
questions  of
regularisation, measure,  preservation of conformal invariance,  etc   should
be properly taken
into account. These issues are  easy   to resolve at the semiclassical level
\wit\  but
they become  quite  subtle  once  one  tries  to  obtain exact  results. A  way
 to  by-pass
these   complications  \tsw\ is to  find first  the {\it effective action}
$\G_{gwzw}$  in the
\Gwzw  and  then  identify it with the {\it effective action}  $\G_{sm}$  of
the corresponding
sigma model.  Let  $S(g,A,\g)$ be the classical action  of a \Gwzw defined on a
curved  $2d$
background ($g$ is an element of a group $G$, $A_m$ is the $2d$ gauge field
taking  values
in the algebra of  a subgroup $H$, and $\g_{mn}$ is a $2d$ metric).
The  quantum effective action $\G_{gwzw} (g,A,\g)$ for the fields  $g,A$  is
given by
$$ \G_{gwzw} = S(g,A,\g) + ({\rm quantum \ \ corrections }) $$ $$   =
S(g,A,\g) + \int
d^2z \sqrt
\g \ R  \varphi (A,g) + ... \ .  \eq{1.1} $$
  Let  $S(x^\m, \g)$ be the classical  action of a  \sm
which should   correspond   to the \Gwzw
$$S(x, \g) = {1\over { 4 \pi \a' }} \int d^2 z \sqrt {\g} \ [  \del_m x^\m
\del^m x^\n G_{\m
\n}(x)\  + \ i \ep^{mn} \del_m x^\m \del_n x^\n  B_{\m\n}(x) \  +  \ \a'  R  \p
(x)
\ ] \  \ .
\eq{1.2} $$
Here $x^\m$ ($\m = 1, ...,$ dim$G/H $)  are some    coordinates on $G/H$.
The  quantum effective action in the theory (1.2) has the following symbolic
form
$$ \G_{sm} = S (x, \g) +  ({\rm nonlocal \ \  terms})  \ .  \eq{1.3} $$
The arguments in (1.1) and (1.3) are already classical (background) fields.
The idea is to find the \sm action $S(x,\g)$ by  comparing  the effective
actions (1.1) and (1.3).
That means one should  solve for the gauge field in (1.1), fix a gauge  and
then  identify
the {\it local}  second-derivative part of the result  as $S(x, \g)$.
It is clear that in  deriving the \sm action one can  ignore all possible
non-local terms which may appear in $\G_{gwzw}(g,A, \g)$ or in the process of
solving for the
gauge field.\foot {The issue of  non-local terms  was a matter of  some
confusion   in \tsw. }
The derivation of the effective action  $\G_{gwzw}$  which is the basic element
of this
approach  \tsw\ will be further clarified and extended to the supersymmetric
case below.

We shall begin (Sec.2) with a  discussion of the operator approach. After a
review
 of refs.\dvv\bsfet\  we shall point out  that
the   answers to the questions   why the resulting   background metric  is  a
`deformed' one
(i.e. is  different from the standard $G$-invariant metric on $G/H$) and why
one  gets
  a non-constant  dilaton   with    $\sqrt G \ {\rm e}^{-2\p }$  being
$k$-independent are
closely related. There exists a close connection with the discussion    \mor\
of  the analogy
between the structure of Hamiltonians of quasi-exactly-solvable  quantum
mechanics systems
\tur\ and that of the stress tensor of conformal theories based on generalised
\hal\mor\  Sugawara or affine-Virasoro
construction. The scalar $\phi$ which  was interpreted   as an `imaginary
phase' in \mor\
is, in fact, the dilaton  of the  \sm corresponding to a given conformal
theory. As a result,
we show  in general that  the dilaton   which appeares in the operator approach
 is given by the
logarithm of the ratio of the determinants of the \sm metric and an invariant
metric on $G/H$,
so that the  combination $\sqrt G \ {\rm e}^{-2\p }$  does not depend on the
matrix   in the
bilinear form of the currents in the stress energy tensor (in particular, it is
 $k$-independent).
We also  comment on possible application of a similar approach to  providing  a
 \sm
interpretation  to   conformal theories   based on  more  general solutions
 of the `master equation'   constructed in \hal\mor\halp\ (it is clear that the
dilaton
field will in general be non-trivial as in the case of the coset models).

Secs.3,4,5  are devoted to the  description of the field-theoretic approach  to
derivation of the
\sm couplings corresponding to  $G/H$ coset conformal theories. As a
preparation for the analysis
of the \Gwzw  case,    in Sec.3  we   present  the expression for the effective
action in the
ungauged WZW theory \tsw.  We   explicitly include the
effect of field renormalisation which makes the effective action non-local. We
compare our result
($ S(g)=kI(g) \ra  \G (g) = (k + \half c_G) I(g') $) with other `effective
actions' in WZW theory
which appeared  in the literature \polles\sch\boer\ger.   In Sec.3.2  we
generalise
the  analysis to
the case of the $N=1$ supersymmetric WZW theory emphasising that  there is no
shift of $k$ in the
resulting effective action  $\G (g) = k  I(g')$  (the shifts of $k$ due to
fermionic and bosonic
contributions  cancel each other).

In Sec.4.1 we    consider
the derivation  of the effective action in the bosonic gauged WZW theory.   The
quantisation of  gauged $G/H$ WZW theory \gwz\karabali\  is based on
representing the
corresponding  path integral  in terms of  the path integrals in ungauged  WZW
theories for   the
group and subgroup.   This makes it possible to use the  analysis  of the
ungauged  WZW theory
carried out in Sec.3. We clarify the
  previous discussion  in \tsw\  by  pointing out  that  for the purpose of
deriving the
corresponding local \sm action the   non-local terms in $\G$  introduced by
field renormalisations
can be ignored.   For the same reason, it is possible to ignore the non-local
terms which are
of cubic and higher order in the gauge potential.  As a result, we  obtain a
`truncated'
effective action $\G_{tr} (g, A)$ (4.17),(4.18) which is  quadratic in  $A_m$
but still
non-local. We compare our approach with that of \bs\ (where  a dimensionally
reduced  $d=1$
form of the effective action was used)  explaining that while the  \sm metric
 and dilaton we should get should
be equivalent to that of \bs\  our direct approach makes possible also to
compute  the
antisymmetric tensor coupling.\foot {  The general expression for the
antisymmetric tensor
coupling we shall get is equivalent to the expression  already  suggested in
\bs\   using the
analogy with the
 result for the metric.   Ref. \bs\ contains also the derivation (without
assuming the $1d$ reduction) of the antisymmetric tensor in a   particular case
of the
  $SL(2,R)\times SO(1,1)/SO(1,1) $  ($D=3$ ``black string")  model. }

 In analogy with the case of the ungauged \SWZW  (Sec.3.2)
the derivation of the effective action in  gauged \SWZW  in Sec.4.2
can be effectively reduced to the discussion of the bosonic case.
As in the bosonic case and the case of ungauged \SWZW  our treatment  of the
gauged \SWZW  is  in  correspondence  with
the results of the operator approach to the superconformal coset theory
\gko\fuch\ks.  We use manifestly supersymmetric approach  which is parallel
to the one  of ref. \karabali\ in the bosonic case  with the fields replaced by
superfields
 (our approach is different from  the previous  path integral analysis  of this
theory in \schn).  We find that up  to the non-local corrections
introduced by the field renormalisations
the effective action  of gauged \SWZW is equal to the {\it classical}   action
of the  bosonic
gauged WZW theory.  We note that the absence of a shift of $k$ is consistent
with perturbation
theory.  As a consequence,  the exact form in  the  corresponding \sm will be
the same as the
`semiclassical' form of the \sm in  the bosonic theory. This conclusion is  in
agreement  with  the one  obtained  in the operator approach in \Jack\ (in the
case of the
$SL(2,R)/U(1)$  supersymmetric theory) and in \bsfet\ (in the case of a
general $G/H$
supersymmetric theory).

In Sec.5  we  start with the `truncated' effective action of the \Gwzw and
eliminate
 the gauge field  (Sec.5.1).  The  local part of the resulting action  is then
put in the
\sm form (Sec.5.2)  and the corresponding metric, antisymmetric tensor and
dilaton couplings are
determined  (the  general  form  of the  \sm couplings  we find   is equivalent
 to the
expressions  presented  in \bs). The background fields   are
non-trivial functions of the parameter
 $ b =    { c_H- c_G\over 2(k+ \half c_G) }=\fourth  (c_H- c_G)\a'$ and
describe  a
large class of
conformal  sigma models.   In  Sec.5.3  we put the metric into a more explicit
form by  making the key observation that  as a consequence
 of the gauge invariance  (before  gauge fixing) the metric defined on the
full group space   has  dim$H$  null vectors.   We  represent the metric    in
terms of
a particular  basis orthogonal to the null vectors and compute its inverse and
determinant.
The metric can be considered as a deformation of a `standard' invariant metric
on the coset space
$G/H$.

Finally, in Sec.6  we establish the equivalence between the results for the
metric and dilaton
found in the operator and field-theoretic approaches. In particular, we
explicitly prove that
dilaton  and the metric found in Sec.5   satisfy relation
$ \sqrt G \ {\rm e}^{-2 \p}= \sqrt {G^{(0)}}$  where $G^{(0)}_{\m\n}$   is
 the standard metric on $G/H$ (and is $k$-independent).

\newsec { Operator Approach to Derivation of Background Geometry Corresponding
to Coset Conformal
Theories}
 \subsec {Basic  ideas }
A  possible strategy of
determining the  geometry  corresponding to a given conformal theory
is to try to
interpret the Virasoro condition $ (L_0 + {\bar L}_0 - 2) F  = 0$  on states as
 linear field
equations  in some background and
 to extract the expressions for the background fields from the explicit form of
the
differential operators involved.  The  marginal operators $F$ of
conformal theory serve
as `probes' of geometry, so that one may be able to  determine   the
corresponding
metric, etc.  from their
equations just as from  geodesic equations or field equations in a curved
space.

In order to implement this program one is  to make a number of important
assumptions.
First, one should  specify  which configuration (`target') space $M$  (with
coordinates $x^\m
, \ \m = 1,...,D$) should be used,  so that  $F$ will   be parametrised   by
fields on $M$, and $L_0$
acting on $F$  will reduce to differential operators  on $M$. Next, one  should
 understand how to
represent
 the resulting equations in terms of background fields.
 The  main   assumption is
that  the
conformal theory should correspond to a \sm
$$S = {1\over { 4 \pi \a' }} \int d^2 z \sqrt {\g} \ [      \del_m x^\m
\del^m x^\n G_{\m
\n}(x)    +  i \ep^{mn} \del_m x^\m \del_n x^\n  B_{\m\n}(x)    $$ $$  +  \a'
R  \p (x)
 +   T (x)  +  ...  ]  \  .
\eq{2.1} $$
If this assumption is true  (for example, if  a  Lagrangian formulation of a
conformal theory is
known  and the  existence of a \sm  representation can be  checked in the
`semiclassical'
approximation) then the `anomalous dimension  operator',
i.e. the  derivative of the $\b$-functions at the conformal point $ ({\del
\b^i\ov \varphi^j})_*$
should  be equivalent to the `Klein-Gordon' operator $L_0 + {\bar L}_0$ for the
 corresponding
marginal  perturbations of conformal theory.   One is  thus to  invoke the
knowledge of the structure
of the \sm   conformal  anomaly  coefficients (`$\b$-functions'), or the
effective action which
generates them $$  S = \int d^{D}x \sqrt {\mathstrut G} e^{- 2 \phi} \lbrace {2
\over 3} (D-26) -
\alpha^{\prime} [R + 4 (\partial_{\mu} \phi)^{2} - {1 \over 12} H^{2}_{\lambda
\mu \nu}]
$$ $$  + {1 \over 16} [\alpha^{\prime} (\partial_{\mu} T)^{2} - 4T^{2}] +
\ldots
\rbrace \ .  \eq{2.2} $$
One should  start with this background-independent action, linearise the
corresponding equations
near an arbitrary  background, and compare them  with the equations for the
corresponding states
in conformal theory.  The  equations for the tachyon,  graviton, dilaton   and
the
antisymmetric tensor perturbations
 ($T , \ h= G-G_* , \ \varphi = \p - \p_* , \ b = B- B_* $; in what
follows we shall omit
the superscript $*$ indicating background fields) take the following   symbolic
form  ($\a' = 1$)
$$  ( - \D + 2 G^{\m\n} \dm \p \dn)T  - 4T + ... =0     \ , \ \
\ \ \ \D \equiv  {1 \over  {\sqrt G} }
\del_\m (  {\sqrt G} G^{\m\n} \del_\n )\ ,\eq{2.3} $$
$$  ( - \D + 2 G^{\m\n} \dm \p \dn)h + Rh + H\del b  + ... = 0 \ , \eq{2.4} $$
$$ ( - \D + 2 G^{\m\n} \dm \p \dn)\varphi   +  H\del b  + R\del^2 h +  ... = 0
\ ,
 \eq{2.5} $$
$$  ( - \D + 2 G^{\m\n} \dm \p \dn)b +  H\del h + ... = 0 \ . \eq{2.6} $$
Given  a  second order
differential equation  which follows from the $L_0$-condition
for the lowest scalar `tachyonic' state
it  should  be
possible to determine the  corresponding  background metric and dilaton
 by looking at the coefficients of the terms which are second and first  order
in derivatives
   and comparing them with (2.3).   To
determine the antisymmetric tensor field strength one  should   compare the
first-derivative terms
in the  equations for `massless' perturbations with  the corresponding terms in
(2.4),(2.5),(2.6).

It should be emphasised   that if this approach works at all,
 its   consistency
should not be  surprising.  If the correspondence  between a  conformal \sm
and a conformal
field theory
exists  in a given case, then    solutions  of the conformal invariance
conditions that follow from
(2.2)   should represent  a    conformal point; their perturbations   should
correspond to
marginal perturbations of a conformal theory, i.e. the  equations for the
latter {\it
must} have  the form of
(2.3)--(2.6).

\subsec { $SL(2,R)/U(1)$ model }
 This   `operator' approach was used in  \dvv\  to determine the exact
metric and dilaton
backgrounds corresponding to the  $SL(2,R)/U(1)$ model (see also \Jack\ for
the
supersymmetric case)
and   was  later applied to  more general   $G/H$ coset models in \bsfet.
 Related   ideas were  discussed in \anton\mor.  Let us
first briefly recall
the argument from \dvv.  The stress tensor of the $\sl$ model  can be
represented in the form
$$ T_{zz}  = {1 \over k-2 } \eta^{AB} J_A(z) J_B(z) - {1\over k}  (J_3(z))^2 \
,   $$
where $ \eta_{AB}$ is the metric on the Lie algebra of $SL(2,R)$.  It is
sufficient to consider
only the zero modes. The equations for the physical scalar state (tachyon)
$T(g)= T(r, \tl , \tR )
$  ($r, \tl , \tR $  are coordinates on $SL(2,R), \    g= e^{{i\over 2 } \t_L
\s_2 } e^{\ha
r\s_1} e^{{i\over 2 } \t_R \s_2  } $) are
$$ (L_0 + {\bar L}_0 - 2) T = 0 \ \ , \
\eq{2.7} $$ $$ \  \   (L_0 - {\bar L}_0 ) T = 0 \ \ . \  \eq{2.8} $$
{}From the expressions for the  zero modes of the left (and right) currents (we
shall use the
same notation $J_A$ for the zero modes of the currents $J_A(z)$)  $$ J_{\pm} =
\e{\pm i \tl} [
{\del \over\del r} \pm i (\sinh r )^{-1} ({\del \over\del\tR} - \cosh r {\del
\over\del\tl}) ] \ ,
\ \ \ J_3 = i {\del \over\del\tl} $$
 one finds that \dvv
$$ L_0 = - {1 \over k-2 }\Delta_0 - {1\over k} {\del^2 \over\del\tl^2} \ , \ \
\
 {\bar  L}_0 = - {1 \over k-2 }\Delta_0 - {1\over k} {\del^2 \over\del\tR^2} \
, $$
$$ \D_0 = {\del^2 \over\del r^2} + \coth r {\del \over \del r }
+ (\sinh r )^{-2}  ( {\del^2 \over\del\tl^2} - 2 \cosh r {\del^2 \over\del\tR
\del \tl}
 + {\del^2 \over\del\tR^2} ) \ ,  $$
so that (2.8) is satisfied if
$$ T= T(r, \t)  + {\tilde T} ( r, \tt) \ \ , \ \  \ \t \equiv  \ha (\tl -\tR)
\ \ ,
\ \ \ \tt \equiv  \ha (\tl +\tR)  \ \ . $$
Restricting $L_0$ to  $T(r, \t)$ we get
$$ L_0 = - {1 \over k-2 }[{\del^2 \over\del r^2} + \coth r {\del \over \del r }
+ (\coth^2 {r\over 2} - {2\over k} )   {\del^2 \over\del\t }]  \ .  $$
As a result, eq.(2.7) can be represented as a covariant   Laplace equation
$$ 2 (L_0 -1) T(r,\t) =     [ - {1 \over \e{-2 \p} {\sqrt G} } \del_\m ( \e{-2
\p} {\sqrt G}
G^{\m\n} \del_\n ) -
 2 ] T = 0 \ , \eq{2.9} $$
where $x^\m = (r, \t) $ and
$$ G_{\m\n} dx^\m dx^\n= \ha (k-2) [ dr^2 + f (r) d \t^2 ]\ , \ \
\ f(r)={ 4\th^2 {r\over 2} \over
 1-{2\over k} \th^2 {r\over 2}}\ ,  \eq{2.10} $$
$$ \p = \p_0  - \ha  \ln  \sh r
+ \fourth \ln  f (r)  \ . \eq{2.11} $$
The metric (2.10)  can be rewritten  also as follows
$$ G_{\m\n} dx^\m dx^\n= \ha (k-2) [ dr^2 +  \e{4(\p -\p_0)} {\rm sinh}^2 r  d
\t^2 ]\ .
$$ Hence the  deformation  of the metric from the canonical metric on the
homogeneous space
$SL(2,R)/U(1)$  (or the standard metric on $S^2$ in the case of the
$SU(2)/U(1)$ model where $r=
i \varphi $)  can be attributed to the presence of a non-constant dilaton
background.
One can also give an opposite interpretation: the fact that the metric which
is extracted from
the $L_0$-operator of conformal theory turns out to be different from the
invariant metric on
the coset space implies the presence of a non-constant dilaton background.
Note also that  the
``measure" combination $$   \sqrt G \ \e{-2 \p} = a \ \sh r \ \ ,
 \ \ \ a= \ha \e{-2 \p_0 }  (k-2) = \const  \  $$
is proportional to the  canonical measure on the coset  and thus is essentially
independent of
$k$ (one can  make $a=1$   by  an  appropriate choice of the constant $\p_0$).
All these
observations  are not accidental and will be given a  systematic explanation
below.

The leading-order form of the expansion of the metric (2.10) and the dilaton
(2.11) in powers of
$1/k$ solves the one-loop  Weyl invariance conditions for the corresponding
$D=2$ sigma model
\efr\wit. To satisfy the  ``beta-function" equations  at higher loop orders one
must  include
corrections to the metric and the dilaton which (up to a field redefinition)
are in a agreement
with  the exact representation (2.10),(2.11)  as was checked  up to three- and
four-loop orders in \ts\ and \Jack.   A similar argument in the case of
supersymmetric
$\sl$ model
(where $1/(k-2)$ in $T_{zz}$ is replaced  by $1/k$ and thus there is no  $2/k $
term in the
brackets in $L_0$)   shows that there are no $1/ k$ corrections to the
leading-order metric and
the dilaton  of the bosonic Euclidean black-hole  background \Jack\bsfet\ (this
is
again consistent
with the sigma model perturbation theory up to five loops \Jack).

\subsec { General expressions for the metric and dilaton in the case  of
affine-Virasoro construction}
It is straightforward to  give a formal generalisation of    the above analysis
to the case of an
arbitrary $G/H$ coset model
  (see \bsfet\sak ).  Let $T_A = (T_a , T_i )$ be the
generators of $G$  where $T_a$ are the generators of $H$  ($A= 1,..., D_G ; \
a=1,..., D_H; \ i =
1,  ..., D; \ D= D_G - D_H $).  Let us  consider again the equations
(2.7),(2.8) for
the lowest level physical $H$-invariant  scalar   state  $T(x^\m)$,   where
$x^\m$
($\m = 1, ..., D $) are
coordinates on $G/H$ ($D $   combinations of coordinates $x^M$ on $G$ which are
invariant under $H$).  Now  the zero mode part of $L_0$ is given by \gko
$$ L_0 = \ha \big( {1 \over \kg } \eta^{AB} J_AJ_B  - {1\over \kh}  \eta^{ab}
J_a J_b  \big) \ ,
 $$
where $$ f^{ACD} f^B_{\ CD} = c_G \eta^{AB}\ , \ \ \  f^{acd} f^b_{\ cd} = c_H
\eta^{ab}$$
 and $ J^A_0$ and  $ J^a_0$ are the zero modes of the
 (left) currents corresponding to the group and the subgroup.
 Representing the currents in  terms of differential operators
$$ J_A =  E_A^{M} (x) {\del \ov \del x^M} \  , \ \ \
J_a =  E_a^{M} (x) {\del \ov \del x^M} \  ,  $$
and considering only  $H$-invariant states, i.e. $(J_a - {\bar J}_a) T =0 \ , $
we can identify
$L_0-1$ acting on $T(x^\m)$ with the covariant Laplace operator (2.9).
Then the   background metric and the dilaton are determined by
 (we restrict all the fields to depend  on $x^\m$ only)
$$ G^{\m\n}= -({1 \over \kg }E^{A\m} E_A^\n  - {1 \over \kh }E^{a\m} E_a^\n )\
\ , \eq{2.12} $$
$$ \del_\m \ln (\sqrt G \ \e{-2 \p } ) =
- \Gmn ( {1 \over \kg }E^{A\l}\del_\l E_A^\n  - {1 \over \kh }E^{a\l}\del_\l
E_a^\n )\ \ .
 $$
This metric and the dilaton  were  computed explicitly for a number of models
(in particular, for $SO(D-1,2)/SO(D-1,1)$  with $  D=3,4$) in \bsfet\ and it
was  observed that
as in the $D=2$ case  the combination $\sqrt G \ {\rm e}^{-2 \p}$ is
$k$-independent.

One is  naturally  led to  the following questions:  why is the resulting
metric
different from the standard $G$-invariant metric on $G/H$,  why  do we get
  a non-constant  dilaton  and  why  is  $\sqrt G \ {\rm e}^{-2\p }$
$k$-independent?
    These questions  turn out to be closely related.
Our approach  is partly inspired by   the discussion \mor\ of the
analogy between the
structure of Hamiltonians of quasi-exactly-solvable (QES)  quantum-mechanical
systems \tur\ and that
of the stress tensor of conformal theories based on  affine-Virasoro
construction
\hal\mor.
 Given a compact finite-dimensional (simple)  Lie group $G$ and a set of its
generators $J_{ A}$,  the Hamiltonian of the QES system can be put in the form
(we
shall ignore
possible terms which are linear in $J_A$) $$ \H =   \C^{AB} J_A J_B     \ \ , \
\eq{2.13} $$
where $\C_{AB}$ is a constant symmetric matrix.  At the same time, the
holomorphic stress tensor
of a conformal theory based on generalised  Sugawara construction \hal\mor\
is
represented  by
$$ T_{zz}  = \C^{AB} J_A(z) J_B(z)     \ \ , \ \eq{2.14} $$
where  $J^A (z) $  are
the generators of an affine (Kac-Moody) algebra   determined by the structure
constants
$f^A_{\ BC}$ and the Killing metric $\eta_{AB}$ of $G$  (with the central term
proportional to
$\k_{AB}= k \eta_{AB}$). The condition that $T_{zz}$ should satisfy the
Virasoro
algebra imposes a
 `master equation'   on  $\C^{AB}$ \hal\mor
$$ \C^{AB} = 2 \C^{AC} \k_{CD}\C^{DB} - \C^{CD}\C^{KL} f_{CK}^A f_{DL}^B -
\C^{CD}f_{CL}^K f_{DK}^{(A} \C^{B)L} \  \eq{2.15} $$
(the central charge of the Virasoro algebra is $C= 2 \k_{AB} \C^{AB}$).
The standard Sugawara-GKO solution of (2.15) is
$$ \C^{AB}=  {1 \over \kg } \eta^{AB}  - {1\over \kh}  \eta^{AB}_H   \ ,
\eq{2.16} $$
where $\eta^{AB}_H $ denotes the projector on the Lie algebra of $H$.

As  is  clear  from the above discussion, the background  metric
corresponding to a  conformal
theory based on the stress tensor (2.14)  can be determined by looking only at
the zero mode part
of (2.14), i.e. at  its  `dimensionally reduced'  analogue (2.13).  As was
discussed
in \tur\mor, if
the generators $J_A$ of $G$ can be realised as vector fields on a manifold $M$
(which  will  play the
role of a configuration space of a quantum mechanical system)  and are
anti-Hermitian with respect
to a scalar product defined by some metric $\go$ on $M$ then the operator
(2.13)  reduces to a
covariant Laplacian  on $M$  with the  metric determined by $\cab$
and $\go$  and an
extra scalar field (the scalar field term can be traded for a potential by a
phase
 transformation).  In the case when the Hamiltonian (2.13) originates from  a
conformal theory   $\cab$  is not arbitrary but  is  a solution of the  `master
equation'
(2.15). Also, the choice of the configuration space $M$ is implicitly dictated
by  the conformal
theory.

In general, unless  we consider particular solutions of (2.15)
 which may have non-trivial extra symmetries  (commuting operators),  the only
natural choice for a configuration space is the group space $G$  itself.
Representing the zero
modes of the currents  $J^A (z)$ and $\J^A (z)$  as differential operators  on
$G$
(with coordinates $x^M$)
$$  J_A=  E_A^M (x) \del_M \ , \ \ \ \J_A=  \E_A^M (x) \del_M \ , \eq{2.17} $$
$$ G_{0MN} =\eta_{AB} E^A_M E^B_{N}= \eta_{AB} \E^A_M \E^B_{N} \ , \ \ \
\eq{2.18} $$ $$
[J_A,J_B] = f^C_{AB} J_C \ , \ \ [\J_A,\J_B] = f^C_{BA}\J_C\ , \ \
[J_A,\J_B] = 0 \ ,  \eq{2.19} $$
 ($E_A^M$ and $E_A^M$ are the left-invariant  and right-invariant  vielbeins on
$G$;
 indices are raised and lowered with $\eta_{AB}$ and $ G_{0MN}$)
 we get   from  the zero mode part of $ (L_0 + {\bar L}_0 - 2) F  = 0$ the
following equation for the
lowest scalar
 state  $T(x)$
$$  [ - ( G^{MN} \del_M\del_N   +  G^N \del_N)  - 2] T (x) = 0 \ , \eq{2.20} $$
$$ G^{MN} =  -\ha \C^{AB} (E^M_A E^N_B + \E^M_A \E^N_B) \ , \ \ \
  G^N  = -\ha
    \C^{AB} (E^M_A \del_M E^N_B + \E^M_A \del_M \E^N_B) \ . \eq{2.21} $$
Eq.(2.20) becomes equivalent to  the \sm equation (2.3) if
there exists   a  scalar $\phi$ such that
$$ G^N = G^{MN} \del_M \ln (\sqrt G \ \e{-2 \p } ) \ . $$
In fact,
such $\p$  can be  found explicitly by using  the  obvious properties of
$E^A_M$, or,  equivalently,
  by observing that
that since   $J^A$ and $\J^A$   are   anti-Hermitian
with
respect  to the  invariant scalar product on the group defined by $G_{0MN}$,
  $\ (f,g) = \int d^Dx {\sqrt{
G_{0}}} f^*(x) g(x) $.  One has
$$ \del_M E_A^M = - E_A^M \ \del_M  \ln {\sqrt {G_{0}}}\ , \ \ \ \
 E^A_N\del_M E^M_A = - E^{M}_A\del_M E^A_N\ ,  $$ $$ \ \ \
\del_M E^A_N - \del_N E^A_M = f^A_{BC} E^B_ME^C_N \ .  $$
As a consequence,
$$ \p =  \ha \ln  {\sqrt {G \ov G_{0}} }  \ , \  \ \
{\rm i.e.}
 \ \ \ \ \sqrt G \ \e{-2 \p} ={\sqrt {G_{0}}} \ . \eq{2.22} $$
The `measure factor' in (2.22) is  thus    universal, i.e. is independent  of
$\C^{AB}$.
To understand a  simple origin of this result   one should  compare the term in
the effective
action (2.2) leading to (2.3)
 $$ \int  d^{D_G} x{\sqrt G} e^{-2 \p } G^{MN} \del_M T \del_N T$$
with the
`expectation value' of the zero-mode `Hamiltonian' $$ (T, \H
T)=
\int d^{D_G} x {\sqrt{
G_{0}}} T(x) \H  T(x) \ , \ \ \ \  \H =\ha \C^{AB} (J_A J_B + \J_A \J_B) $$
  and use the antihermiticity of  the currents with respect to the left-right
symmetric measure
defined by $G_0$.
 The dilaton's role is to compensate for the fact that the two scalar products
have {\it different} measures.

 The dilaton
field is  {\it  non-trivial}  because  in general the metric $G_{MN}$ is not
equal to  the
canonical Killing metric $G_{0MN}$ on $G$.\foot {Note that the determinant of
$G^{MN}$ in (2.21) can not be factorised into the product
of the determinant of $\C^{AB}$ and the rest because $\E^M_A$ is different
from $E^M_A$.}   The dilaton is constant only in the
simplest  case of the standard Sugawara  solution  $\C^{AB} \sim \eta^{AB} $
corresponding to  the group $G$.   A `deformation' of the metric is directly
related to the
non-triviality  of $\C^{AB}$ which is  dictated by  the conformal invariance
(Virasoro) condition
(2.15). If there exists  the  corresponding Lorentz-invariant \sm it should
also contain the
antisymmetric tensor   coupling.\foot {For  an attempt to construct a
field-theoretic realisation of the  affine-Virasoro construction see \haly,
where, in particular,  the presence of  the  antisymmetric tensor coupling  was
 pointed out.  One may  expect that
it may be possible to reinterpret   the `master equation' (2.15)  as the
(one-loop) Weyl
invariance condition  ${\hat R}_{MN} + 2 D_M D_N \p =0 $  for the metric
$G_{MN}$, antisymmetric
tensor  {\it and} the dilaton
 (the operator product relation from which  (2.16) is
derived probably corresponds to using (a non-local) renormalisation  scheme  in
which only a
one-loop contribution is present in the sigma model Weyl anomaly coefficients;
  the condition of Weyl invariance at higher loop orders will be automatically
satisfied as a consequence  of the 1-loop relation  {\it and} the  Kac-Moody
algebra).
 This is to be compared with the approach of \haly\ where (2.16) was
interpreted as an
Einstein-like equation on the group space but the dilaton  was  not
introduced.}.

There seems to emerge  an  interesting  connection between algebraic   and
geometric aspects of
such conformal theories  (and corresponding string   solutions).  The  geometry
is
determined by  a  choice of the group {\it and}  a  choice of  a particular
solution  of the
`master equation'.
 The  question about a relation between group-theoretic and
geometric aspects  of a  similar   construction was raised independently  in
the
quantum mechanical
context in \mor\tur\  and   in \haly. Now we see that  once the condition  of
conformal invariance
 is satisfied, the geometry  which   appears  is
that of
the corresponding string solutions described by conformal sigma models.

\subsec {Explicit expressions for the  (inverse)  metric  and dilaton in the
case of the
$G/H$ coset conformal theory }
Let us now return to the $G/H$ case, i.e. specialise  the  general expressions
(2.21),(2.22) to the
solution  of  (2.15) corresponding to the $G/H$ coset conformal theory  with
$\C^{AB}$ given
by (2.17).  Here  the main assumption of an  existence of the \sm description
is
satisfied  in view of
 the existence of the Lagrangian formulation
 in terms of gauged WZW models (this assumption
can be checked  explicitly, e.g.,  in the semiclassical approximation).
In this case  there is  an extra  symmetry  which makes it possible
to   subject the states to the `$H$-invariance' condition $(J_a   - \J_a) F
=0$.
In particular,
$$  (J_a  - \J_a) T  =0 \ , \ \ \   Z^M_a \del_M T = 0 \ , \ \ \ Z^M_a \equiv
E^M_a - \E^M_a
\ . \eq{2.23} $$
As a result, $T$ can be restricted to depend only on  $D=D_G - D_H$ coordinates
$x^\m$ of the coset
space $G/H$ which will thus play the role of the configuration space of the
corresponding sigma model.
The presence of the constraint (2.23) implies that  the metric we will get
from
(2.20),(2.21)
will be  a  `projected' one.
Let us   define
 the projection operator on the subspace orthogonal (with respect to $G_{0MN}$)
to $Z^M_a$
$$ \Pi^N_M
\equiv  \delta^N_M -  Z^N_a (ZZ)^{-1ab} Z_{Mb} \ , \ \ \ \ (Z
Z)_{ab} = G_{0MN} Z^M_a Z^N_{b}\ , \ \ \ \  \Pi^2 = \Pi \ . \eq{2.24} $$
Then
$$  G^{MN}=  \Pi^M_K {\hat G}^{KL} \Pi_L^N \ ,   \ \ \
{\hat G}^{MN}=  {1 \over \kg } \eta^{AB}E^M_AE^N_B   - {1\over \kh}  \eta^{ab}
E^M_aE^N_b \ ,
\eq{2.25} $$  $$ {\hat G}^{MN}={1
 \over \kg } ( E^M_AE^{AN} -  \g   E^M_aE^{aN}) = {1 \over \kg }[
E^M_iE^{iN}
-   (\g -1)   E^M_aE^{aN}] \ ,  \eq{2.26}   $$ $$
 \g =  {k+\ha c_G\over  k+\ha c_H } \ , \ \ \ \   \g - 1 ={ c_G - c_H\over 2
(k+\ha c_H) } \ .   \eq{2.27} $$
We have split  the  indices $A= (a,i) , \ i = 1,...,D$ on the indices
corresponding to the
subalgebra   and  the indices corresponding to the tangent space to $G/H$.  If
one  solves
(2.23) explicitly,  replacing  $x^M$  by  the  coset space coordinates $x^\m$,
which are some
$D$ invariant combinations of $x^M$  such that
 $$    Z^M_a  H_M^\m = 0 \ , \ \ \  H_M^\m =   {\del x^\m \ov \del x^M } \ ,
\eq{2.28} $$
then  the metric (2.25)   takes the form
 $$  G^{\m \n }=  H^\m_M G^{MN}H^\n_N =  H^\m_M {\hat G}^{MN}H^\n_N \ , \ \ \
 $$ $$  G^{\m \n }= {1
\over \kg } ( E^{\m A} E^\n_A   - \g    E^{\m a}E^\n_a ) \ , \  \ \ E^\m_A
\equiv H^\m_M E^M_A
. \eq{2.29} $$
In the simplest case $H^\m_M = \delta^\m_M$. More generally,
one can choose any set of vectors $H^\m_M$ which are orthogonal to $Z^M_a$.
 As a result, we  will
 get again  eqs. (2.20)--(2.22)  with the tensor indices $M,N,...,$  restricted
 to
 $G/H$, i.e.
replaced by $\m,\n, ...$.
 Since in the present case (under the constraint (2.23))
the  operators $J_A$ are  anti-Hermitian
with respect to  the invariant metric on the coset
$$ \go_{\m\n} = \eta_{ij} E^i_\m E^j_\n \ , \eq{2.30}  $$
 we find as in (2.22)
  $$ \p =  \ha \ln  {\sqrt {G \ov \go } }\ .  \eq{2.31} $$
Similar expression for the scalar $\p$  was  given  in \mor\foot{
It was  assumed in \mor\  that
the generators $J_A$   can be   realised as  vector fields on the  homogeneous
space $M= G/H$.
In the context of quantum mechanical applications, the choice of $H$ need not
be   related to the
choice of $\C^{AB}$. However, from the  point of view of determining the
geometrical
background corresponding to a coset conformal field theory   the possibility to
realise the generators $J_A$ as vector fields on $G/H$ is related to the fact
that   it is
sufficient to restrict consideration to $H$-invariant states.
  A one-dimensional quantum mechanical system with the configuration space $M$
can be
 considered at the same time as  a dimensional reduction of  a $2d$ sigma model
corresponding to
the  conformal theory.}
 where  $\p$ was interpreted in as an
`imaginary phase' since one can replace  the  scalar term  $G^{\m \n} \dm \p
\dn $  by  a potential,
 performing  a
similarity transformation.\foot {
 Then $\H = \e{\p } \H' \e{ -\p }  , \  \H' = - \D + V  ,  \  \ V= G^{\m\n}
\dm \p \dn \p - \D \p \ . $
It is interesting to note
that when  $\Gmn$ and
$\p$ satisfy the  sigma model Weyl invariance conditions the leading order term
in the potential
$V$  is  equal to $\fourth R$ (plus a constant central charge deficit).}

The  above expressions (2.29)--(2.31)  lead us  again to the following {
conclusions}:

1. the metric $\Gmn$ is  different from the standard $G$-invariant metric
$G^{(0)}_{\m\n}$ on $G/H$
because  the matrix $\C^{AB}$, in general, is different from the Killing metric
$\eta^{AB}$
of  $G$;

2. the presence of a non-trivial dilaton is a consequence of $\Gmn \not=
G^{(0)}_{\m\n}$, i.e. of
$\C^{AB}\not= \const \ \eta^{AB}$;

3.  $ \sqrt G \ {\rm e}^{-2 \p}$ is equal to the $G$-invariant
measure factor
 ${\sqrt {G^{(0)}}}$ on
$M=G/H$   and thus   automatically   does not depend on $\C^{AB}$.  In
particular,
it
does not depend on the parameter $k$ of the  coset  conformal theory (2.17).
This provides a
simple explanation of the fact of  $k$-independence of  $ \sqrt G \ {\rm e}^{-2
\p}$
anticipated  (on the basis of  explicit examples and path integral measure
considerations) in
\kir\bsft.

It should   certainly be possible  to  compute   also the antisymmetric tensor
background  by
comparing the equation for a massless $(1,1)$ state with (2.4)--(2.6).   We
shall
find the
antisymmetric tensor  and also reproduce the expressions for the metric (2.29)
and  the dilaton
(2.31) in  Secs.5,6  by using  the direct   field-theoretical approach
starting with  the gauged
WZW theory (which provides a Lagrangian
formulation of the coset conformal  theory).

Let us mention that one can also use the operator approach
 to determine the background geometry  in the case of superconformal coset
models.
For example,  one can give  a sigma
model interpretation to the $N=2$ Kazama-Suzuki models \ks\ based on K\"ahler
$G/H$ spaces. The
corresponding  metrics  will be different from the  invariant K\"ahler metrics
on $G/H$
 and the dilaton will be  non-trivial  (see  \bsfet).

It  is  clear, at the same time,  that the operator approach  has a number of
obvious
shortcomings.  It is indirect and  based on a number of implicit assumptions.
{\it If } a given conformal theory admits a Lagrangian formulation (and there
exists a weak coupling
limit) then it  should be possible to derive the  corresponding  exact \sm
action
using field-theoretical methods. This will be  demonstrated    below  for   the
case of $G/H$ model.

\newsec {Effective Action in  WZW Theory  }
\subsec {Bosonic  WZW theory}
  Below we shall first   review the argument  which leads to the expression
for the effective action in WZW theory  suggested in \tsw\ and then   comment
on relations to
other approaches.
 Up to a field renormalisation, our effective action is
essentially  equal to the classical WZW action with the shifted $k, \ k\ra
\kg$
($c_G$ = the eigenvalue of the  second  Casimir  operator in the adjoint
representation).  From one point of view, the shift is related to the Legendre
transformation
involved; from another,  it is   a one-loop phenomenon originating from a
determinant
(cf.\shifley\shif).

The WZW theory \witt\nov\ is defined by the action
 $$ S=
{k }  I(g) \  ,  \ \  \    I \equiv  {1\over 2\pi }
\int d^2 z  \Tr (\del g^{-1}
\bd g )  +  {i\over  12 \pi   } \int d^3 z \Tr ( g^{-1} dg)^3   \ \ . \
\eq{3.1}
$$
As it is easy to see,  using the Polyakov-Wiegmann identity \polwig,
$$ I(ab) = I(a) + I(b) - {1\over \pi }  \int d^2z \Tr (a^{-1} \del a \ \bd b
b^{-1} ) \ \ ,  \eq{3.2} $$
the generating functional $W(B)$ of the correlators of the currents
$$
  \ee{-W(B) } =  \int [dg] \  \e{ - S(g)  +  B\bar j (g) } \ \ ,  \ \eq{3.3} $$
 $$ B{\bar j} (g) \equiv   {k\over \pi }  \int d^2z \Tr (B \J ) \ ,
\ \ \ \ \J \equiv \bd g g^{-1} \ , \eq{3.4} $$
 is given by \polwig\polles
 $$   W(B) = - {k  }  I(u) \  , \ \
\  \ \ B = u^{-1} \del u  \ \ .  \eq{3.5} $$
$W$ does not receive quantum
corrections being   equal to  the classical action  evaluated on the
classical solution depending on $B$ \shifley.

We  would like to determine  the quantum effective action  $\G $  for the
original
chiral field $g$ itself. As discussed in \tsw\ it can be represented as a
`quantum' Legendre
transform of $W(B)$
$$   \ee{- \G (g) } =  \int [dB] \  \e{ - W(B)  +  B\bar j (g) }
 =
\int [dg'] \  \e{ - S(g') }  \ \delta [  \J (g') - \J  (g) ]
 \ \ .  \  \eq{3.6} $$
 To compute
(3.6)  we change the variable from   $ B = u^{-1} \del u  $ to $u$ and  define
the resulting
Jacobian in the left-right symmetric way as the square root of the non-chiral
determinant
\polwig\gwzw\oal
 $$
 \det ( \del + [B, \ ] )  \det ( \bd + [\bar B , \ ] )  =
 \exp { [   c_G I(v^{-1}u) ] } \det \del \det \bd \ \ ,
\eq{3.7} $$
$$  \ B= u^{-1} \del u \
\ , \ \ \ \
 \  \bar B = v^{-1} \bd v   \ \ $$
with  $\B =0$, i.e.
 $$ [d B ]  = [du]\  ``{\rm det} ( \del + [B,\  ] )"  = [du] \
\exp { [ \ha c_G I (u) ] } \ ({\rm det}\del\bd )^{1/2}
 \ \  .  \ \eq{3.8} $$
The extra $\ha$ in front of $c_G$  in (3.8) which is absent in the standard
expression  for the chiral determinant
is very important,  being a way of  implementing  the  left-right symmetry of
the theory. It is
necessary in order to get the correct shift of $k$ in the final expression for
the effective
action.
  We find
 $$ \ee{- \G (g) } =  N \int [du] \  \exp { [ (k + \half c_G) I(u)   +  B(u)
\bar j (g) ]} \ \ .  \  \eq{3.9} $$
This integral is computed  by the same  method  as (3.3), i.e. by   using
the `non-abelian generalisation of the gaussian integral' \polwig\ (see
(3.2),(3.4),(3.5))
and we get   $$  \G (g) = (k + \half c_G) I(g')  \ \ , \eq{3.10} $$
where the field renormalization  is understood in the following sense
$$ \bd g' g'^{-1}=  {k \over k + \ha c_G } \bd g g^{-1} . \eq{3.11} $$
 Eq.(3.11)   corresponds to the renormalisation of the current
$ k \bar J = (k + \half c_G){\bar J}'$.

This action has the right symmetries (conformal and chiral $G\times G$
invariance) one would like
to preserve at the quantum level.\foot {  It may be possible to prove that
these symmetries fix
$\G$ uniquely up to the two  ($k$- and current-) renormalisation constants
using, e.g., the `quantum action principle' and BRST cohomology  method, cf.
\bep.}
Note that because of the field renormalisation  (3.11) the effective action
becomes non-local
when expressed in terms  of the original field $g$.  Using the
parametrisation
$\bd g g^{-1}= T_A \E^A_M (x)  \bd x^M \ $  ($T_A$ are the generators of $G$)
one can solve
(3.11)  within   the  $1/k$ - expansion,
$$ \ x'^M =  x^M + \sum_{n=1}^{\infty}
{1\ov k^n}
y_n^M(x) \ .
$$ The corrections $y_n^M$   contain inverse  powers of differential
operators, i.e.
 are non-local functionals of $x^M(z). $ For example,
 $$ y_1^M = - \ha c_G
P^M_K  \bd x^K \  , \ \
\ \
\  {(P^{-1})}^M_K\equiv
\d^M_K \bd  + \E^M_A \del_K \E^A_N \bd x^N\ .  $$

 The  functional $\G (g)  $ (3.6)  should be   equivalent to the  standard
 generating functional of 1-PI  correlators  of the field $g$ itself.
  Although  this  is not obvious,      the  resulting action (3.10)
 is perfectly consistent  with the presence of the shifted $k$
in the quantum    equations of motion    and the  stress  tensor
in the operator approach  to  WZW model as   conformal  theory \kniz
$$ (k + \half c_G) \del g (z,\bar z) = \ :J_A(z)  g (z,\bar z)T^A:\ , \ \ \ \
 (k + \half c_G) \bd g (z,\bar z) = \ :{\bar J}_A(\bar z) T^A g (z,\bar z) :\ ,
$$ $$
 T_{zz} =  {1 \over \kg } \eta^{AB}:J_A(z)J_B(z): \  . $$
The action (3.10) can be considered as a `classical' representation of these
quantum relations with
the normal ordering suppressed.\foot {
Note that the
background value of the stress tensor  should be given by the variation of the
effective
 action over the $2d$ metric. As follows from (3.10),(3.11), $ T_{{\bar z}
{\bar z}} = ( \kg )
(\bd g' g'^{-1 })^2 =  {1 \over k + \half c_G } (k\bd g g^{-1})^2 $.
The  field renormalisation in (3.10) is
actually  a renormalisation of the currents, equivalent to the one in the
quantum equations of
motion. }

Alternatively, one  may start with  the  assumption
 that the  effective  action $\G (g) $ in the WZW theory must satisfy conformal
and chiral
$G\times G$ invariance conditions. Then a  natural  (and probably unique)
choice for such
$\G$ is  the classical action itself, up to  the renormalisations of $k$ and
the  current,
$\G (g) = k' I (g') , \  \bd g' g'^{-1}=  Z   \bd g g^{-1}$.  Correspondence
with the  c.f.t.
approach then   fixes $k'= k + \ha c_G$ (and probably fixes also  $Z={k \over k
+ \ha c_G }$).
  A possibility to  find   an  exact expression for the effective  action of
the WZW theory
should   not be  surprising, given its solubility in the operator approach.

 Computed on a curved $2d$ background
the quantum effective action  contains also  the usual Weyl anomaly term (see
e.g. \shifley)\foot
{ There  is  no extra term which is a non-trivial functional of both $g$ and
the $2d$
metric $\g_{mn}$  since   the  vanishing of the $\b$-function of the WZW model
implies that the
operator of the trace of the stress tensor is proportional to $C R $.} $$
\G_{anom} (\g) =   - {C
\ov   96 \pi }  \int R {\D}^{-1} R \ , \ \ \ \
  C = {k D_G \ov\kg}     \ .          \eq{3.12}  $$
As is well known \polwig, the  WZW action (3.1) can  also  be   represented
as
a non-local functional of the current
$$  I(g) = \o  ( J ) \  , \ \ \ \   J =  g^{-1 }\del g   \ , \eq{3.13} $$
$$  \o (J)=
- \1p \in \Tr \{ J \sum_{n=0}^\infty  {(-1)^{n} \ov n+2 }([{1\over \del}J ,\ \
])^n  {\bd\over
\del}J  \}$$ $$ = - \1p \in \Tr \{ \half J{\bd\over \del} J  - \third J[{1\over
\del}J , {\bd\over
\del}J ] + ... \} \   ,\eq{3.14} $$
 or,
$$  I(g) = \W  (\J ) \  , \ \ \ \   \J = \bd g g^{-1 }  \ , \eq{3.15} $$
$$  \W(\J)
=- \1p \in \Tr \{ \J \sum_{n=0}^\infty {1\ov n+2 } ([{1\over \bd}\J ,\ \ ])^n
{\del\over \bd}\J
\} $$ $$ = -
\1p \in \Tr \{\half \J{\del\over \bd} \J  + \third \J[{1\over \bd}\J ,
{\del\over \bd}\J ] + ... \}
\  .\eq{3.16} $$
This suggests an alternative way of computing  the path integral (3.6).
Changing  first
the quantum variable  from $g'$ to $\J'$ and taking into account the
contribution of the
Jacobian  as in  (3.8),(3.9) we get
$$   \ee{- \G (g) }  =
\int [d\J'] \  \exp \{ - (k+ \ha c_G) \W(\J')  \}  \ \delta [  \J' - \J  (g) ]
 \ \ .  \  \eq{3.17} $$
The formal  use  of the $\d$-function gives
$$ \G (g) =  (k+ \ha c_G) \W(\J) =(k+ \ha c_G) I(g) \ , \eq{3.18} $$
i.e. we reproduce (3.10) but {\it without}  a  field renormalisation.  The
reason for  an apparent
paradox is that a regularisation prescription implicit  in the
formalism  based on eqs. (3.2),(3.7) \polwig\gwzw\oal\  is consistent  with the
formal
$\d$-function identities for the composite  currents  only  if an extra field
renormalisation  is
included (see in this connection  \kutas\dev\boer).
A related paradox  is found if one differentiates  (3.6) over the 2-metric.
Since the classical stress tensor $\bar T $ is  proportional to $ {1\ov k}
\J^2$,
 naively using the  $\delta$-function identity one finds the same expression
(with unshifted $k$) for the derivative of $\G$.  The correct of the
$\delta$-function identity
in the present context is: $< F [\J (g')]  \delta [  \J (g') - \J  (g) ] > =
F [ Z\J (g)],
\ Z= \sqrt {k\ov k+ \ha c_G}.$  As we noted already, such an identity is try
only under a
particular choice of the regularisation scheme.

Maintaining  equivalence  between the local field theory and   operator
conformal theory
results is rather subtle and depends on a choice of a particular regularisation
prescription
(which should correspond to a normal ordering prescription in c.f.t.).
As in the case of  the $3d$ Chern-Simons theory \wittt\  the  one-loop shift of
$k$
in the effective action may happen in one regularisation and not happen in
another one
(see e.g. \shif).  The absence of a renormalisation of $k$ in the   standard
Legendre transform
of the generating functional for  correlators of currents (which does not
receive loop
corrections \shifley) and its presence in the `quantum' Legendre transform (34)
seems  related to
an observation  \ver\ that  similar `quantum' Legendre transform in $SL(2,R)$
Chern-Simons
theory  relates  two representations (in terms of  affine and Virasoro
conformal blocks) with
`bare' and  renormalised values of $k$.

 Let us now compare (3.10) with other `effective actions' in WZW theory  which
appeared  in the
literature.
Similar effective actions were  recently discussed  in the context  of
`induced' gauge theory (and  $2d$ gravity)    \polles\zam\sch\boer.
Since the   fermionic  determinants  in a background gauge field are  expressed
in terms of
 $\o (A) $ or $ \W (\A) $  (with the coefficient $-k$, cf.(3.7))
 one  can take   the `induced'  action $ S (A) = - k \o (A) $ as  a
classical
action and    find   the corresponding  quantum effective action for $A$.
Introducing the
source $\J $ for $A$  one can compute  the   generating functional $W(\J)$  by
integrating
  $\exp ( - S(A) + \J A ) $ over $ A $ (cf. (3.3)).   Using (3.2) and
the standard expression \polwig\ for the determinant (3.8) (without $\ha$ in
front of $c_G$) one
obtains: $ W (\J) =  - (k+ c_G) I (u) , \ \J = - (k+ c_G)\bd u u^{-1} $.  If
the effective action
$S_{eff} (A) $ is  defined as  the Legendre transform of $W(\J )$,  then\foot {
In general,   the
Legendre  transform of the functional $ W(A) = a  \o ( b A )= a I(g)  ,\  bA =
g\inv \del g   $ is
given by
 $  {\tilde W }  ( {\bar B}) =  a \o ( b A ) - {1\ov \pi } \int d^2z {\bar B}A
=
- a \W ( c{\bar B}) = -a I ( u) \ , \ \  c B = \bd u u^{-1} \  , \ \ c =
(ab)^{-1} $
(the coefficients are easily checked in the abelian case). }
  $$
S_{eff} (A)=   - (k+
c_G) \o (A) \ . \eq{3.19}   $$
 To  have  a consistency with the  one-loop perturbative computation
of $S_{eff}$
\polles\sch\ one needs also to include  a  field renormalisation factor
  so that
the  conjecture for the   exact  form of the  effective action  reads
\polles\sch\boer
 $$S_{eff} (A)=   - (k+ c_G) \ \o (Z  A) \    , \ \ \ Z  = 1 -  {c_G
\ov 2k } + O({1\ov k^2}) \ .
 \eq{3.19'} $$
 A non-trivial expression for $Z$
reflects  a
  choice  of a  specific  regularisation prescription  made  in perturbative
computations (see
also  the  above remarks concerning (3.18)).  It is not clear, however, that
this choice
is consistent with the conformal field theory  approach
since straightforward application of the Polyakov-Wiegmann identity and
the expression \polwig\ for the chiral determinant  (which is  derived using
Ward identity)
gives (3.19) with $Z=1$.

The   expression (3.19), though similar to  our action (3.10),(3.11) in
structure,  is
different in two respects.  The   minus sign in (3.19)  is due to the  fact
that  one has
started with the  induced  WZW action (with coefficient $-k$). The   factor of
$\ha$
difference in the shifts of $k$ in the overall coefficients in  (3.10) and in
(3.19)
is related to the fact that (3.10) is an  effective action (in the
left-right symmetric WZW theory) for the field $g$ itself     (cf.(3.6))
 while (3.19)  was  derived  using   the  ``chiral" field  $A$  as    the main
quantum
variable.

One may   also draw an analogy between the  quantum effective action (3.10)
and  the  free field action  which   was suggested in \ger\ as a basis for the
free field
formulation of the WZW conformal theory.  Using the Gauss decomposition  in
upper triangular, diagonal and lower triangular
matrices     $ g = g_U (\psi) g_D (\vp ) g_L (\hi) $   to parametrise $g$ in
terms of the
fields  $ \hi_\a$ and $\ps_\a$  (labelled by all positive roots $\a$ of the
algebra of  $G$)
and the field $\vp_n$ (taking values in the Cartan torus, $n= 1, ...,r ; \  r =
{\rm rank}\ G$)
one finds for the classical action  (3.1) \ger
 $$  S = {k\over 2\pi }
\int d^2 z  [ \Tr (\del g_D^{-1}
\bd g_D )  +  2 \Tr
(J' \bd g_L g_L^{-1} )] \ ,  \ \ \ J' (\ps , \vp ) \equiv  (g_U g_D)^{-1} \del
(g_U g_D)\ .
   $$
Introducing  instead of $\ps_\a$ the new field $W_\a$ such that
$$ k \Tr [ J' (\ps , \vp )  (\bd g_L {g_L}^{-1})(\hi)] = W_\a \bd \hi_\a  \ ,
\eq{3.20} $$
one can represent (3.12) in the form \ger
$$ S = {1\over \pi }
\int d^2 z (    W_\a \bd \hi_\a      -   \ha    k \del \vp_n \bd \vp_n  )   \ .
 \eq{3.21}  $$
The change of variables $ \ps_\a \ra W_\a = Y_{\a}^\b (\hi , \vp , \ps ) \del
\ps_\b $
is accompanied by a Jacobian. The latter should be defined in such a way   that
  the
resulting  free theory is consistent with the operator approach to WZW
conformal model
based  the affine  algebra and Sugawara representation.  In particular, the
Jacobian should depend
only on $\vp_n$ and the $2d$ metric $\g$ \ger.  Its logarithm  contains  three
terms: $   c_G
\del \vp_n \bd \vp_n  $   (which  leads to a shift of  the value of the
coefficient $k$ in
(3.21));  $ \ \r_n \vp_n R $ ($ \r_n  = $  one half of the sum of all positive
roots); and a
pure Weyl anomaly  term.
 The final `quantum WZW action' on a curved $2d$ background then takes the form
(in the conformal gauge) \ger
 $$ S_q
= {1\over \pi } \int d^2 z\  [    W_\a \bd \hi_\a      -  \ha    (\kg ) \del
\vp_n \bd \vp_n
   -  \fourth \r_n \vp_n {\sqrt \g}R     ] $$ $$
 - \  {(D_G -r)  \ov   192 \pi }  \int R \ {\D}^{-1}
R       \ .  \eq{3.22}  $$
The shift of $k$ in (3.22) is the same as in (3.10) (one can of course rescale
$W_\a$
to make the shifted $k$  appearing  in front of the whole action) and,  as in
(3.10),  originates
from a 1-loop determinant. Note  that  the
fields in (3.15) are still quantum; in particular,   they need to be integrated
 out to get the
correct  central charge term (3.12) \ger : $$ C = \ha {(D_G -r) }  +  r  -  {
12 \r^2 \ov \kg }   +
\ha   {(D_G -r) } =  {k D_G \ov \kg }\ . $$
The  background  values of the fields $W_\a ,
\hi_\a , \vp_n $
  should correspond to  the argument $g$ of the effective action (3.10).


\subsec { Supersymmetric WZW theory }
Computation of the effective action in  supersymmetric WZW theory
can be reduced to that  in the bosonic WZW theory by observing
\div\sus\suss\ that  by a  formal field redefinition the action of
the supersymmetric WZW
theory can be
represented   as a  sum  of the bosonic WZW action and the
action of  the free Majorana fermions in the adjoint representation of the
group $G$.
As was  noted in \suss\red\schn,\  the transformation of fermions  which is
needed to decouple
 them from $g$ is {\it chiral}  (the interaction term is $\bar \psi (1+\g_5)
\g^a
\del_a g g^{-1} \psi
$) and  therefore   produces a non-trivial Jacobian. The logarithm of the
fermionic determinant
gives a contribution  proportional to the  bosonic  WZW action. The net result
is  the
shift of the coefficient $k$ in the bosonic part of the action \red :
$$ k I (g, \psi ) \ra   {\hat k} I(g)  +  I_0 ( \psi) \ , \ \ \ \ \ \ {\hat
k}\equiv k- \ha c_G  \ .
\eq{3.23} $$ This  gives, in particular,  the following expression for the
central charge
\red\fuch\schn $$  C_{susy}  =  C ({\hat k} )  + \ha D_G =
 { k- \ha c_G  \ov k} D_G + \ha D_G =   ( {3\ov 2}  - {c_G  \ov 2k } )D_G \ ,
\eq{3.24} $$
implying that the central charge  contains only   the leading $1\ov k
$-correction (i.e.
the two-loop  correction  in the  perturbative  expansion). The expression
(3.24) is consistent with
the absence of the  3- and 4-loop corrections  to the dilatonic $\b$-function
in the corresponding
\sm \all.

Note that though the  question about the shift of $k$  in (3.23) may look
ambiguous,  the correct choice is actually fixed by  the condition of
correspondence with
the perturbation
theory.
 Had one started  with the often-used component form of the supersymmetric WZW
action in which   bosons and fermions are decoupled from the  very beginning,
one  would get no
shift of $k$ in (3.23).   A    prescription  in which there  is   no shift of
$k$ (and
hence the central charge, is given by the naive expression \nem\ $C_{susy} =
C(G,k)  + \ha D_G$
 containing  terms  of all orders in $1/k$)  is    inconsistent  with the
standard
renormalisation scheme  employed in perturbative \sm computations (such a
prescription is
effectively non-local when  considered from the \sm point of view).\foot { As
discussed in \all,\
a consistency of  the  shift of $k$ in (3.23) and hence of (3.24)  with
expectations about higher
loop ($L\geq 5$) corrections to the dilaton $\b$-function  presumably  relies
on  a  choice of a
specific regularisation scheme  (in which a version of the    Adler-Bardeen
theorem is  true).   }\foot {  The shift of $k$  is a  direct consequence of a
manifestly supersymmetric  approach. Once one makes a chiral rotation to
decouple fermions from
bosons,   the new supersymmetric transformation laws will involve $\gamma_5$
 and  probably will be anomalous at higher loop orders.}
 The  quantum equivalence
of the supersymmetric WZW theory to the bosonic WZW theory and the set of
free fermions was proved also at the level of the full  conformal field theory
in the operator
approach \fuch\ (in particular, it was shown that, in agreement with
(3.23),(3.24),
  the super
Kac-Moody algebra
 with  the central parameter  $k$ reduces   to the direct product of the
bosonic Kac-Moody algebra
with the parameter ${\hat k}= k- \ha c_G $ and  the  free fermionic algebra).

To find the
   (bosonic part of the )    effective action corresponding to  the \SWZW   one
needs   to
repeat  the argument which led to  the expression (3.10),(3.11) for the
effective action in the
bosonic WZW theory.  Instead of (3.3) we get
$$
  \ee{-W(B) } =  \int [dg][d\ps ]  \  \exp \{ - kI(g, \ps )  +  B\bar j (g) \}
$$ $$ =
     N \int [dg] \  \exp \{ - (k- \ha c_G)I(g)  +  B\bar j (g)  \} \ \ .  \
\eq{3.25}
$$         Though  the source  term  in (3.25)
  contains the original unshifted parameter $k$ this does not  matter  at the
end since $B$ is
integrated out in  $\G$ (3.6). As a result, the  effective action in the
supersymmetric WZW theory
is obtained by replacing  $k$   by  ${\hat k}= k- \ha c_G$  in (3.10),(3.11),
$$\G (g) = k  I(g')  \  , \ \ \ \        \eq{3.26} $$
$$ \bd g' g'^{-1}=  (1-   { c_G \ov 2k }) \bd g g^{-1} , \eq{3.27} $$
i.e. it is equal to the classical WZW action with {\it unshifted}  $k$.
The  shift of $k$   in $\G$  produced by integrating  out fermions
 in (3.25)    is exactly cancelled
 out by the  contribution  of the bosonic determinant (3.8) in (3.9).
Note also that the field renormalisation  relation (3.27) contains  only a
one-loop correction.


\newsec {Effective Action in  gauged WZW Theory  }
\subsec {Bosonic   gauged WZW theory}
In this section we shall first  consider
a  derivation  of the effective action in the bosonic gauged WZW theory
(clarifying the
discussion in \tsw)    and then  generalise it to the supersymmetric gauged
WZW theory case. The
quantisation of the   gauged $G/H$ WZW theory \gwz\karabali\  is based on
representing the
corresponding  path integral  in terms of   path integrals in ungauged  WZW
theories for   the
group and the subgroup.   That  makes it possible to use the  analysis  of the
ungauged  WZW theory
carried out in the previous section.

 The classical  gauged  WZW action \witt\gwzw\
 $$ I(g,A) = I(g)  +{1\over \pi }
 \int d^2 z \Tr \bigl( - A\,\bd g g\inv +
 \bar A \,g\inv\del g + g\inv A g \bar A  - A \A \bigr)$$ $$ \equiv I_0(g,A) -
\1p \int
d^2z \Tr ( A\A )  \   \eq{4.1} $$
is invariant under  the  standard  vector $H$ -  gauge transformations ($A, \A$
take values in the
algebra of $H$)
 $$ g \ra u^{-1} g u \ , \  \ A \ra u\inv ( A + \del ) u \ , \ \
 \A \ra u\inv ( \A + \bd ) u \ , \ \  \ \ u = u (z, \bar z)  \ .$$
The two terms in (4.1) $I_0$ and $\int \Tr (A\A)$ are separately invariant
under
the $holomorphic$ vector gauge transformations $ g \ra
u^{-1}(z) g u(\bar z) $.

 Parametrising  $A$ and $\A$ in terms of $h$ and $\bh$  which
take values in  $H$
 and transform as $ h \ra u\inv h \ , \ \ \bh \ra   u\inv \bh $,
$$ A = h \del h\inv \ \ , \ \ \ \A = \bh
\bd \bh\inv  \ \ , \eq{4.2} $$
one can use  the  Polyakov-Wiegmann identity  (3.2) to represent the gauged
action   as the
difference of the two gauge-invariant terms: the ungauged  WZW actions
corresponding to the group
$G$
and  subgroup $H$,
$$ I(g,A) = I (h\inv g \bh ) -  I (h\inv \bh)  \ \ . \eq{4.3} $$
Since
 $$ I(h\inv g \bh ) = I_0 (g,A)  + I(h\inv) + I(\bh) \  , \ \ \
  I(h\inv \bh) = \1p \int d^2 z \Tr (A\A)  +   I(h\inv) +
I(\bh) \ \  , $$
 the non-local terms $I(h\inv) + I(\bh)$  cancel
out in the  classical action (4.3) (but will survive in the effective one since
the  coefficients
of the two terms in (4.3) will get different quantum  corrections).
 Changing the variables
in   the path integral
$$ Z = \int [dg] [dA][d\A] \ \ee{ - k I(g,A) }      \ \  \eq{4.4} $$
and  using  the expression (3.7)  for the non-chiral determinant
\polwig\gwzw\oal,
$$ \det (\del + [A, \ ] ) \det (\bd + [ \A , \ ] )
= \exp{ [ c_H I(h\inv \bh ) ]} \det \del \det \bd \ \ , \ \eq{4.5} $$
we get \karabali
$$Z = N\int [dg] [dh][d\bh] \ \exp{ [ - k I (h\inv g \bh ) + (k + c_H)  I
(h\inv
\bh)    ]    }
$$
$$ = N' \int [d\tg] [d\tth] \ \exp [- k I ( \tg) - (-k-c_H)I(\tth)] \ ,
\eq{4.6} $$
$$  \tg\equiv h\inv g \bh\ , \ \ \ \
\tth \equiv  h\inv \bh \ ,  $$
where  a gauge fixing was assumed ($N'$ is proportional to the product of $D_H$
free scalar
determinants originating from the Jacobian of the change of variables).
One concludes
\karabali\ that   $ k I(g,A)$  can be quantised as the  `product'  of the two
WZW theories for the
groups $G$ and $H$ with  the levels $k$ and $-(k+c_H)$, or, equivalently, as
the `ratio' $G_k/H_k$
of the WZW theories with the  equal levels $k$. This is clear, for example,
from the  resulting
expression for central charge [3,4]
 $$ C(G/H) = C(G,k) + C(H,-k-c_H) - 2 D_H = C(G,k) - C(H,k)\ , \eq{4.7}  $$
 $$ \ \ \ C(G,k)
\equiv  {kD_G\ov k+\ha c_G} \ ,  $$
where $-2D_H$ corresponds to the contribution of the free determinants in $N$.

 One possible  approach to derivation of the effective action in the gauge
theory
(4.4) is to introduce sources for all fields $(g, A, \A )$ and to use the
background field method.
However,  we prefer to  couple sources only to gauge-invariant combinations of
fields $\tg$ and
$\tth$ in (4.7).  Then the problem is reduced to   computation of  the
effective actions in  the two
 decoupled ungauged WZW theories.
 According to our discussion  in Sec.3 to  find  the
 effective action  of  the  WZW theory  for group $G$ one is to replace its
parameter $k$ by
$k + \half c_G$ and to renormalise the field  (see (3.10),(3.11)).
This  implies  that the effective
action in the gauged WZW theory  is   given by
 (note that $-(k +c_H ) + \half c_H = - (k+ \half c_H)$)
 $$ \G (g,A) = (k + \half c_G)I (\tg' ) - (k + \half  c_H)  I (\tth')         \
,  \eq{4.8} $$
$$   \bd \tg' \tg'^{-1}=  {k \over k + \ha c_G } \bd \tg \tg^{-1} \ , \ \ \ \
\
  \bd \tth' \tth'^{-1}=  {k + c_H\over k + \ha c_H } \bd \tth \tth^{-1} .
\eq{4.9} $$
As in the ungauged case,  the  structure of  the effective action (4.8) is in
a natural
correspondence with that  of the
 stress   tensor    in the conformal field theory
approach \gko\
$ T_{zz}  = {1 \over k + \half c_G } : J_G^2 :\  - \  {1 \over k + \half c_H }
:
J_H^2 : \ .$

The field renormalisations (4.9)  complicate the problem of representing  the
action (4.8) in terms
of the original fields $g, A$ and $\A$.  However, as explained in the
Introduction,
for the  purpose  of  finding  a  \sm which corresponds to the gauged WZW
theory  it is
sufficient to   consider   only  the {\it local} part of the effective action
(4.8). In what follows
we shall drop out  various non-local parts of (4.8) in several stages. First,
  we  shall  ignore the field renormalisations (4.9) (since they introduce
additional
non-localities, see Sec.3), i.e.  we  shall   replace (4.8) by the following
action  (which is
gauge-invariant and, in general, still non-local)
 $$ \G' (g,A) = (k + \half c_G)I (\tg ) - (k + \half  c_H)  I (\tth)         \
.  \eq{4.10} $$
Using (4.2),(4.3),(4.7)
 we can represent  (4.10)  in terms of $g$
    and the gauge field \tsw
$$ \G' (g, A) = (k + \half c_G) I (g, A) + \ha  ( c_G -   c_H)  \O (A)
\ .  \eq{4.11} $$
Here $\O (A)$  is a non-local  gauge invariant functional of $A $ and $  \A $,
$$ \O (A) \equiv I(h\inv \bh) = \o (A) + \W (-\A) + \1p \int d^2 z \Tr (A\A)
   \ \  , \eq{4.12} $$
where  the functionals $\o$ and $\W$ have already appeared in (3.13)--(3.16)
$$ \ \ \ \o (A)  = I(h\inv) = - \1p \in
\Tr \{ \half A{\bd\over \del} A  - \third A[{1\over \del}A , {\bd\over \del}A]
+  O(A^4)  \}
,\eq{4.13} $$
$$ \W(-\A)  =  I(\bh)
 = -\1p \in \Tr \{\half \A{\del\over \bd} \A  -  \third \A[{1\over \bd}\A,
{\del\over \bd}\A] +
O(\A^4) \} \  . \eq{4.14} $$
The `quantum correction' $\O(A)$  (equivalent to the  induced action
corresponding
to  Dirac fermions)  contains both local and non-local terms.
   As it is clear from  (4.13),(4.14) (see also (3.14),(3.16))  all  terms  in
$\O$
which are $O(A^n,
\A^m)$ with $n,m \geq 3$ are non-local, i.e.
 $$ \G' (g, A) = (k + \half c_G) I (g, A) + \ha  ( c_G -   c_H) \O_0 (A)  +  (
{\rm non-local}  ) \ ,  \eq{4.15} $$
$$  \O_0 (A) \equiv   \1p \in \Tr ( A\A  - \half
A{\bd\over \del} A - \half \A { \del \over \bd}
\A  )  \eq{4.16} $$
$$ =  \2p \int  \Tr  F {1 \over \del \bd } F  \ , \ \ \ \ \ \ \ F \equiv \bd A
- \del \A \ . $$
In  the case  when the subgroup $H$ is abelian,  the higher order non-local
terms in (4.15)
automatically cancel out, i.e.   $\O_0 = \O$ \tsw.   In contrast to the full
$\O$ its   quadratic
part $\O_0$  is invariant only under the abelian  gauge transformations, $A\ra
A + \del \epsilon , \
\A\ra \A + \bd \epsilon $.
Dropping out the   non-local $O(A^n, \A^m)$ terms  in  $\G'$ (i.e. replacing
$\O$ in (4.11) by
$\O_0$)  we reduce it to the following action
$$  \G_{tr}(g, A) = (k + \half c_G) \big[ I (g) +  \D I (g, A)
\big ] \ , \eq{4.17} $$  $$  \D I (g, A)\equiv {1\over \pi }
 \int d^2 z \Tr \big[ (  - A\,\bd g g\inv +
 \bar A \,g\inv\del g + g\inv A g \bar A  -  A \A )  $$
$$ +  \ha  b  \ ( AQA + \A Q^{-1} \A - 2 A\A)   \big]  \ , \eq{4.18} $$
$$    b \equiv  - { (c_G- c_H) \over 2(k+ \half c_G) }  \ ,\ \ \ \ \
 Q\equiv {\bd\over \del} \ , \ \ \ Q^{-1} \equiv {\del \over \bd}  \ .
\eq{4.19} $$
The absence of the full non-abelian gauge invariance of the quantum correction
term  proportional
to $b \O_0$    should   not present a problem,   since it is clear that  the
gauge
invariance    can always be restored by  re-introducing the higher order
non-local terms.

Let us   note that  in terms of    the redefined  gauge  fields  $$
A' = Q^{1/2} A \ , \ \ \ \A' = Q^{-1/2} \A \ ,  \eq{4.20} $$
we can represent  (4.18) in the
form
$$\D I (g, A)= {1\over \pi }
 \int d^2 z \Tr \big[ (- Q^{-1/2}A'\, \bd g g\inv  +
  Q^{1/2} \A'\, g\inv\del g    + g\inv Q^{-1/2}A' g Q^{1/2} \A'  -  A' \A')
$$ $$
+\  \ha  b \  ( A' - \A')^2
  \big] \ .  \eq{4.21} $$
If we were to  naively ignore  the $Q$-insertions,  we  would  obtain  the
following action
(omitting  primes on $A, \A$)
$$  \tilde \G (g, A) = (k + \half c_G) \big[ I (g) +  {\D{\tilde I }} (g, A)
\big ] \ , \eq{4.22} $$
 $$ {\D{\tilde I }} (g, A)=
 {1\over \pi }
 \int d^2 z \Tr \big[ (- A\, \bd g g\inv  +
  \A\, g\inv\del g    + g\inv A g \A  -  A \A)
$$ $$ + \  \ha  b \  ( A - \A)^2  \big] \ .  \eq{4.23} $$
The  dimensionally reduced   form of the  action (4.22),(4.23) was recently
proposed  (for
the same purpose
 of deriving    the couplings of the corresponding  sigma model)  in \bs.  The
main idea in \bs\ was to concentrate only on  the zero mode dynamics, i.e.  to
consider
 a reduction  of the effective action (4.11)  to one dimension taking  all the
fields to
depend on the time  coordinate only  (which  is effectively equivalent to
setting
$\del = \bd$,
i.e. $Q=1$). It was   observed
 that the resulting $1d$ action (which is  invariant under the  $1d$ gauge
transformation)
becomes   local and quadratic in the gauge field.   While this ansatz  makes
possible  to determine
the metric (and dilaton) of the corresponding \sm (since the original  gauged
WZW theory is
Lorentz invariant,
the
 \sm  action  should be given by  $\int d^2z \ \Gmn \del_mx^\m \del^m  x^\n +
... $  so that to find
the target space metric  it is sufficient to compute the term in the effective
action which is
quadratic in the time derivatives of the \sm fields),  it  has  an obvious
deficiency. After all,
we are dealing with a $2d$ theory which, in general,  contains more couplings
than its dimensionally
reduced analogue. In particular,  the  $1d$ ansatz
 does not allow one  to compute the antisymmetric tensor coupling   in a
systematic way.

The truncated effective action (4.17) will be our starting point in the
subsequent  derivation of
the \sm couplings.  To  extract a  local part of (4.18) is somewhat
non-trivial. It is
not correct  to omit  the terms with the operator $Q$ insertions  (since  $Q$
has  dimension zero and
since this  would break  the  gauge invariance of (4.17)); it is  also not
correct  just to replace
$Q$ by 1  (this would break the Lorentz invariance).  As we shall see in Sec.5,
one should
first integrate out the gauge fields  and  {\it then}   discard  all the
non-local terms.
As    is clear  from the above  discussion, our result for the \sm metric
and dilaton should  be the same as in the $1d$ approach of \bs.


\subsec { Supersymmetric  gauged WZW theory }
 As in the case of the ungauged \SWZW
considered  in Sec.3.2 the derivation of the effective action in  gauged \SWZW
can be effectively reduced to the discussion of the bosonic case.
The path integral quantisation  of the supersymmetric theory follows the same
pattern as
in  the previous subsection.
As in the bosonic case and the case of ungauged supersymmetric WZW theory
(Sec.3.2) our treatment
of the gauged \SWZW  will be in  correspondence  with
the results of the operator approach to the superconformal coset theory
\gko\fuch\ks.  To guarantee
the $N=1$ supersymmetry we shall use the superfield formulation  of the theory
as a starting point
(see e.g. \noj).    Our approach  is parallel
to the one  followed  in   \karabali\ in the bosonic case  (with the fields
replaced by
superfields).\foot {
 Our approach  is different from  the previous  path integral analysis  of
gauged \SWZW
 in \schn\ where   component  formalism  was  used and only the case of $G=H$
was considered.
  It
was mentioned \schn, however,   that
 using  a  superfield in place of  the gauge field  one  should be able to
give    a
manifestly supersymmetric  treatment  of the problem. }

The supersymmetric  version of the ungauged WZW action  (3.1)
is obtained by
 replacing $g$ by the
 corresponding superfield and making other standard replacements ($z^a\ra (z^a,
\t , \bar \t ) $,
$ \ \del \ra D$, etc ) \div
$$\hat g= \exp (T_A X^A) \ , \  \ \ \ X^A= x^A + { \t }{ \psi}^A + {\bar \t
}{\bar \psi}^A  + \bar \t \t f^A \ , \ \ \ D= {\del\ov \del \t}- \t {\del\ov
\del z}\ , \eq{4.24} $$
  $$ \hat S=
{k }  \hat I(\hg) \  ,  \ \  \   \  I(\hg)  \equiv  {1\over 2\pi }
\int d^2 z d^2\t   \Tr (D \hg^{-1}
\bar D \hg )  + \  (WZ-term)  \ . \eq{4.25} $$
The  Polyakov-Wiegmann identity  (3.2)  also  has a straightforward
supersymmetric
generalisation. The supersymmetric version of the gauged WZW action (4.1) is
given by
$$ \hat I (\hg, \hat A) = \hat I(\hg )  +{1\over \pi }
 \int d^2 z d^2 \t  \Tr \bigl( - \hat A\,\bar D \hg \hg\inv +
 {\hat {\bar A}} \,\hg\inv D \hg + \hg\inv \hat A \hg {\hat {\bar A}}
 - \hat A {\hat {\bar A }} \bigr)\   \eq{4.26} $$
$$ = \hat I(\tilde \hg) - \hat I(\tilde {\hat h}) \ , \eq{4.27} $$
where  the  gauge superfields $\hat A , \hat { \A}$   take values in the
algebra of  the subgroup
$H$  and (cf. (4.2),(4.3),(4.7))
$$ \hat A = \hat h D {\hat h}\inv \ \ , \ \ \ \hat \A = \hat{ \bh
}\bar D  {\hat \bh }\inv  \ \ , \eq{4.28} $$
$$
\tilde \hg\equiv {\hat h}\inv \hg \hat{\bh}\ , \ \ \ \
\tilde {\hat h} \equiv {\hat h}\inv \hat \bh \ . \eq{4.29} $$
In view of (4.27) the quantisation of the theory can be reduced to that of the
two
ungauged supersymmetric WZW theories corresponding to the group and the
subgroup (cf.(4.6),(4.7)),
$$ Z = \int [d\hg] [d\hat A][d\hat {\A}] \ \exp \{ - k \hat I(\hg , \hat A) \}
  $$
$$ =  \int [d\tilde {\hg }] [d\tilde {\hat h}]  \ {\cal J}  \ \exp \{  - k I
(\tilde {\hat g} ) +
   k  I (\tilde {\hat h})    \} \ \ . \eq{4.30} $$
Here $\cal J $ stands for  the product of Jacobians of the change of
superfield variables
from  $\hat A$ to  $\hat h$  and from $ \hat {\A} $ to $ \hat {\bh} $  (and
includes also a gauge
fixing factor). While in the bosonic case  the  corresponding product
(regularised in the left-right
symmetric way as in  (3.7))  is non-trivial and leads to the shift  of the
coefficient of the
$H$-term in the action  (see (4.6)),  in the superfield case  each of the
Jacobians is   proportional
to  a field-independent factor.  This happens because    the   non-trivial
contribution of the
bosonic determinant  is   cancelled  out by   a  contribution  of the fermionic
one (this
cancellation is similar to that of   the bosonic and fermionic contributions
to  the coefficient
$k$ of  the effective action in the ungauged  supersymmetric WZW theory, see
(3.25),(3.26)). In fact,  as in the bosonic case,
the Jacobian of the change $\hat A \ra \hat h $
can be expressed in terms of the  path integral with the action
$ \int d^2 z d^2 \t  U ( DV + [ \hat A , V])$ , where  $U$ and $V$ are
superfields
of opposite statistics. Re-writing  this action in  component fields and
integrating them out
it is easy to see that this Jacobian is $\hat A$-independent.

The theory  can thus be  represented  as a `product' of the two  supersymmetric
WZW theories for
the  groups  $G$ and $H$   with  levels $k$ and $-k$.
Since we know already the expression (3.26),(3.27) for the effective action in
 the ungauged supersymmetric WZW model,  it is  now easy to write down the
resulting effective
action  in the theory (4.26).  In particular, we conclude that there are no
shifts of $k$.
To  see this in detail at the component level, let us    first
return to the
 component
notation  and make the chiral rotation to decouple fermions from bosons  as
discussed in Sec.3.2.
According to  (3.23),(3.25)  we get the following result
$$ Z =  N' \int [d\tg][d\tth ][d\psi_G]
[d\psi_H]    \  \exp [ - (k- \ha c_G)I(\tg)   + ( k +  \ha c_H)I(\tth)  $$
$$  - I_0(\psi_G) + I_0
(\psi_H) ] \ \ .  \ \eq{4.31} $$
The factor $N'$ contains     determinants of the free fermions
in the adjoint representations of $G$ and $H$  as well as the   contribution
$({\rm det \ } \bd
\del)^{3D_H}$ originating from $\cal J$ in (4.30)  (which provides  the
correct
count of the free degrees of freedom).  Up to these free-theory  factors,  we
can represent the
resulting theory as  the `ratio'  $G_{k-\ha c_G}/H_{k - \ha c_H}$ of the
bosonic WZW theories for
the groups $G$ and $H$   with levels  $k_G= k-\ha c_G$ and  $k_H= k - \ha c_H$
(we separate
 the shift $c_H$  corresponding to the  bosonic change of variable in order to
identify  (4.31)  with
(4.6))
$$ Z =  N' \int [d\tg][d\tth ]   \  \exp \{ -  k_G I(\tg)   + ( k_H +
c_H)I(\tth)
 \} \  .  \
\eq{4.32} $$
In particular,  we get      the   following  expression for the  central
charge
    (cf. (3.24),(4.7))
$$ C_{susy} (G/H)= C(G,k-\ha c_G) + C(H,-k-\ha c_G)  + \ha D_G + \ha D_H - 3D_H
$$
 $$ = [C(G,k-\ha c_G)  + \ha D_G ] - [C(H,-k-\ha c_G) +  \ha  D_H] =
C_{susy}(G) -C_{susy}(H) \ ,
\eq{4.33} $$
$$ C_{susy} (G) =  (1-  {c_G  \ov 2k}) D_G +
\ha D_G\ , \ \ \  \ C_{susy}(H)  =  (1- { c_H  \ov 2k}) D_H + \ha D_H\  $$
(we have included the  contributions of the free fermions  and the factor
contained in $N'$ in
(4.31)).

This  conclusion is  in
agreement  with   the   conformal algebra approach \gko\ks\ (note that in terms
of the shifted level
${\hat k}=k-\ha c_G$    we get the  $G_{{\hat
k}}/H_{{\hat k} + \ha c_G - \ha c_H}$ - theory).
Let us now compare the  above approach with the one based  on starting with
the component
formulation of the gauged \SWZW  in which the fermions are coupled only to $A,
\A$
 (see e.g. \swz\bsf)
$$ Z= \int [dg][dA ][d\A] [d\psi] \ \exp \{- k I(g,A) - I_0(\psi,A) \} \ .
\eq{4.34}
$$
Here $I(g,A)$ is the bosonic action (4.1) and $I_0(\psi,A)$ is the action of
Majorana fermions
(taking values in the orthogonal complement of the algebra of $H$ in the
algebra of $G$) minimally
coupled to $A,\A$.
One can obtain the action in (4.34)  from the classical action (4.26)
by fixing the fermionic part of the gauge invariance  by the gauge condition
$\psi_H =0$
and solving for the fermionic  components of the gauge superfields.
 Integrating over the fermions
and going  through the same  steps as  in  (4.4)--(4.7) we finish with
$$ Z =  N' \int [d\tg][d\tth ]   \  {\exp \{ -  k I(\tg)   +  [ k  +  \ha (c_G
-
c_H) +   c_H] I(\tth)
\} } \ \ \  ,  \ \eq{4.35} $$
where the contribution of fermions is proportional to $c_G-c_H$.  This
expression
 becomes equivalent to (4.31),(4.32)  if $k$ in (4.35)  is replaced by $\hat k
= k - \ha c_G$.
As in the case of the ungauged supersymmetric WZW theory discussed in Sec.3.2,
the  approach based on
the naive  component action  loses  the shift of $k$ and this is inconsistent
with  manifestly
supersymmetric perturbation theory.

Applying the results of  Sec.3.2 and Sec.4.1  we are now ready to write down
the
expression for the
effective action in the gauged supersymmetric WZW theory.  Using either the
representation in terms
of  the  ungauged  supersymmetric WZW theories (4.30)  and  (3.26),(3.27)  or
the equivalent
formulation in terms of the ungauged bosonic WZW theories (4.31) and
(3.10),(3.11) we get the
following expression for the  (bosonic part of)  effective action
 $$   \G_{susy} (g,A) = k I (\tg' ) - k   I (\tth')         \ ,  \eq{4.36} $$
$$   \bd \tg' \tg'^{-1}=  (1-   { c_G \ov 2k })  \bd \tg \tg^{-1} \ , \ \  \
  \bd \tth' \tth'^{-1}=  (1-   { c_H \ov 2k })  \bd \tth \tth^{-1}\ \  .  $$
As in the ungauged \SWZW  but in  contrast  to the result (4.8),(4.9) for  the
bosonic  gauged WZW
theory
 there are  no shifts in the overall coefficients  of the $G$- and $H$- terms
in
$\G_{susy}$.  Ignoring the non-local corrections
introduced by the field renormalisations in (4.36), we  arrive at the
conclusion
 that the local part
of the effective action  of the gauged \SWZW is equal to the {\it classical}
action of the
bosonic  gauged WZW theory
  $$  \G_{susy}^{(loc)} (g, A) = k I(g,A)$$ $$
 = k \big[ I(g)  +{1\over \pi }
 \int d^2 z \Tr \bigl( - A\,\bd g g\inv +
 \bar A \,g\inv\del g + g\inv A g \bar A  - A \A \bigr) \big] \   ,  \eq{4.37}
$$
i.e.,  in contrast to the bosonic case (4.17),(4.18),  it does not contain the
quantum correction  term proportional to $ b = - { (c_G - c_H) \over 2(k+ \half
c_G) }$.

As a consequence,  the exact form of  the  corresponding \sm will be
equivalent to the
`semiclassical' form of the \sm corresponding to the bosonic theory.  This
conclusion is the same as
the one  obtained  in the operator approach in \Jack\ (in the case of the
$SL(2,R)/U(1)$
supersymmetric theory) and in \bsfet\ (in the case of a  general $G/H$
supersymmetric theory).\foot
 {Note that  ref. \bsfet\   used
 ${\hat k}=k-\ha c_G$  instead of $k$.}
  In particular, there is no shift in the overall coefficient $k$.  As
we have already emphasized  above,    it is $k$ and not ${\hat k}$ that is the
coefficient in  both
  classical and effective actions and hence is the parameter that  should be
used in  a
discussion of a correspondence with the bosonic model and  in  a  computation
of the \sm couplings.

One can find the manifestly supersymmetric form of the  corresponding sigma
model
by directly solving for the  gauge superfields  in (4.26).
The result will take the general form of the (1,1) supersymmetric \sm
$ \int d^2z d^2 \t (G_{MN} + B_{MN} ) DX^M \bar DX^N $.
Fixing the gauge  ($X^M \ra X^\mu$) and using component notation it is easy to
read off  the
corresponding connection  with torsion and (from the  quartic fermionic term)
its curvature.
 \newsec { Sigma Model Corresponding   to  Gauged WZW Theory }
\subsec { Elimination of  the gauge field}
In this section we are going to derive the couplings of the  \sm
which   corresponds to  the gauged WZW theory.  Later in  Sec.6  we shall
establish
the equivalence between the  results of the field-theoretical   approach and
the
expressions for the
metric and dilaton  obtained  in the    operator  approach   in Sec.2.   We
shall  also determine
the antisymmetric tensor coupling which is difficult to find in the operator
approach.

 To give a  \sm  interpretation  to a gauged WZW theory one, should,
in principle,  fix  a gauge and  `integrate out'   the gauge field.
 As was explained  in Introduction,  our method  of derivation of the \sm
corresponding to a \Gwzw is  based on  first   finding   the effective action
$\G_{gwzw}$  in the
 gauged WZW theory
  and  then  identifying  it with the effective action for  the  sigma
model. Both effective actions, in principle, contain local as well as non-local
terms.  In order
to   determine  the classical \sm action  (i.e. the local second-derivative
part of $\G_{sm}$) it is
sufficient  to   consider only  the {\it truncated }  part  $\G_{tr}$ of
$\G_{gwzw}$  (4.17),(4.18)
in which some of the non-local terms have  already  been dropped. Since
$\G_{tr}$
is quadratic in the
gauge field   it is possible to treat it  as in the semiclassical
approximation, i.e.
   to  integrate over  the gauge field, to fix  a
gauge, etc. One  may   then  again  ignore   all   non-local
terms which may appear in the process of elimination of  the gauge field.

Let us start with the  following form (4.21) of the truncated effective action
(4.17),(4.18)
$$  \G_{tr} (g, A) = (k + \half c_G) \big[ I (g) +  \D I (g, A)
\big] \ , \eq{5.1} $$
$$\D I (g, A)= {1\over \pi }
 \int d^2 z \Tr \big[ (- A'   \J'  +
 \A' J'      +    g\inv Q^{-1/2}A' g Q^{1/2} \A'  -  A' \A')    $$ $$
+\  \ha  b \  ( A' - \A')^2
  \big] \ .  \eq{5.2} $$
$$ J' =Q^{1/2} (g\inv\del g) \ , \  \ \ \J'=  Q^{-1/2}  (\bd g g\inv )\ , \ \ \
\
A' = Q^{1/2} A \ , \ \ \ \A' = Q^{-1/2} \A  \ ,   $$ $$
\ Q=  {\bd\over \del} \ , \ \ \ \ Q^{-1} =  {\del \over \bd} \ , \ \ \ \
 b =   - { c_G- c_H \over 2(k+ \half c_G) } \ . $$
To integrate over $A, \A$, or, equivalently, over  $A', \A'$ let us first note
that it is possible
to  ignore the factors  $Q^{\pm 1/2}$ in the $O(A\A)$-term in (5.2):  as  it is
easy to
understand,   they  produce only extra non-local terms which we are not
interested in.  Then
(5.2) takes the form $$\D I' (g, A)= {1\over \pi }
 \int d^2 z  \ \big[  L_1 (g, A)  + L_2 ( g, A) \big] \ , \eq{5.3} $$
$$ L_1 = \Tr (- A'   \J'  + \A' J')  \ , \ \ \ \  L_2 =   \Tr \big[ (g\inv A' g
\A'  -  A' \A')
+  \ha  b \  ( A' - \A')^2 \big]\ . $$
In spite  of its form this action  is still Lorentz invariant  since $A',\A'$
in (5.3) are the
redefined fields of (4.20),(4.21).

 It is
useful first to diagonalise $L_2$, the quadratic in $(A,\A )$,   in (5.3).
Let us follow the notation of Sec.2 ($T_A = (T_a , T_i )$ are  the
generators of  the algebra of $G$;  $T_a$ are the generators of  the algebra of
$H$; $\ A=
1,..., D_G ; \ a=1,..., D_H; \ i = 1,  ..., D$) and  define  the
matrices $C_{ab}, \ M_{ab}, \ K_{ab}, \ P_{ab}$ which are functions of $g $
\foot
{In our notation $\eta_{AB}$ is negative definite for a compact group.
Indices $a,b,...$ are raised and lowered with $\eta_{ab}$. We shall sometimes
use 1 to denote  a
unit matrix or $\eta_{ab}$.  }
 $$    C_{ab } = \Tr (T_a g T_b g\inv ) \ , \ \ \ \  \Tr (T_aT_b) = \eta_{ab} \
, \ \ \ \
M_{ab} = C_{ab} - \eta_{ab} \ ,
\eq{5.4} $$
$$  K=  M^{-1} M^T \ , \ \ \ P= \ha ( 1 + K ) \ , \ \ \  \ M_S = \ha (M + M^T)=
MP \ , \ \ \ M^T_{ab}= M_{ba} \ . \eq{5.5} $$
Though $M$  may not be invertible on $H$, it is non-degenerate for a generic
$g$
(at the end we shall  fix a gauge restricting $g$ to $G/H$).
Introducing the following combinations  $B_a, \B_a$ of the  gauge fields $A_a,
\A_a$
( $A =T^a A_a, \ \A =T^a \A_a$)
$$ B= \ha  P^{-1} ( A'-\A') \ , \ \ \ \ \B=\ha  P^{-1} ( KA' + \A') \ ,
\eq{5.6} $$
$$ A'=  \B- B    \ , \ \ \ \ \A'=  \B+ KB    \ ,  $$
we find
$$ L_2 =  A' M \A' + \ha b (A'-\A')^2 = - B \M B  + \B M_S \B  \ , \eq{5.7} $$
$$ \M \equiv   M_S  -  2 b  P^T P    \ , \ \ \ \M^{T} = \M \ .  \eq{5.8}  $$
Note that $B$ is invariant under the abelian gauge transformations,
$A'\ra A' + Q^{1/2} \del \epsilon , \  \A'\ra \A' +   Q^{-1/2}\bd \epsilon $,
i.e. it is
 the analogue  of the transverse part of $A^a_m$ (the $b$-correction term is
contained only in the
$O(B^2)$ part of (5.7); cf. also (5.11) below).  The current term $L_1$ in
(5.3) takes the form
$$ L_1 =   B \cj + \B \tj\ , \ \ \ \ \eq{5.9} $$
$$ \cj_a =  \Tr ( T_a \J'  + T^b K_{ba} J') \ , \ \ \ \
\tj_a = \Tr [ T_a  (  J' -  \J' )]
\ \ . \eq{5.10} $$
Integrating over $B,\B$ we get from (5.7),(5.9)
$$ L_3 (g) =   \fourth \cj \M^{-1} \cj  -
\fourth \tj M_S^{-1} \tj \eq{5.11}  $$ $$ =  \big[ \ \fourth \Tr (T_a \del g
g\inv )
\ (\M^{-1})^{ab}  \Tr (T_b \bd g g\inv ) + \fourth \Tr (T_a   g\inv \del g  ) \
(K
\M^{-1}K^T)^{ab}   \Tr (T_b  g\inv \bd g ) $$ $$  + \ha  \Tr (T_a  g\inv  \del
g  ) \ (K
\M^{-1})^{ab}   \Tr (T_b   \bd g g\inv ) \ \big] $$ $$  - \big[ \ \fourth \Tr
(T_a \del g g\inv )
\ {(M^{-1}_{S})}^{ab}  \Tr (T_b \bd g g\inv )    +  \fourth \Tr (T_a   g\inv
\del g  )
\ {(M^{-1}_{S})}^{ab}   \Tr (T_b  g\inv \bd g )  $$ $$ - \ha \Tr (T_a  g\inv
\del g  ) \
 {(M^{-1}_{S})}^{ab}   \Tr
(T_b   \bd g g\inv ) \ \big] \  + \ (non-local) \ . \eq{5.12} $$
We have used the expressions for the currents $J', \J'$ in  (5.2)  and
rearranged the derivatives
contained in the $Q$-insertions to separate the local part of (5.11) (the
operators $Q^{\pm
1/2}$ either cancel out or  double,    producing $\bd \ov \del$ or $\del \ov
\bd
$ which effectively
interchange  the derivatives $\del$ and $\bd$ in the currents). As as a result,
we  obtain  from
(5.1)  the following   local Lorentz invariant  action
$$ \G_{loc}  (g) =
 {k + \half c_G \over 2\pi }
\int d^2 z \ \{ \big[ \Tr (\del g^{-1}
\bd g )  +  (WZ-term) \big]  $$
$$  +   \big[  \ha  \Tr (T_a \del g g\inv ) \  (\M^{-1}- M^{-1}_{S})^{ab}  \Tr
(T_b \bd g g\inv )
$$ $$ + \ha  \Tr (T_a   g\inv \del g  )\  (K  \M^{-1}K^T- M^{-1}_{S})^{ab}
\Tr (T_b  g\inv \bd g )
$$ $$  +   \Tr (T_a  g\inv  \del g  )\  (K  \M^{-1}+ M^{-1}_{S})^{ab}   \Tr
(T_b \bd g g\inv ) \
\big]\}\  .  \eq{5.13} $$
\subsec {Sigma model representation}
The gauge invariance of the effective action makes it possible to fix a gauge
and
 express  $g$ in terms of $D$ coordinates $x^\m$ on $G/H$.
It is useful, however,  to  start with the full set of $D_G$ coordinates
$x^M$ on the group $G$
space and restrict to $G/H$ only at a later stage. We shall see that  as a
reflection of the gauge invariance of the original action, the resulting metric
 on
$G$  will be   degenerate, having $D_H$ null vectors.
In what follows,    by the \sm fields we shall understand the  expressions
which are found after imposing the gauge condition, i.e. after
replacing  $x^M$  by $x^\m$ and making
  the  corresponding replacements of the
tensor indices, $M,N,...
\ra \m , \n , ...$, etc (see the end of Sec.5.3).

Using the coordinate parametrisation
$$ g\inv \del g = T_A E^A_M (x)  \del x^M \ , \ \ \ \ g\inv \bd g = T_A
E^A_M(x)
 \bd x^M \ ,\ \ \ \   \eq{5.14} $$
$$  \del g  g\inv  = T_A \E^A_M (x)  \del x^M \ , \ \ \ \  \bd g g\inv = T_A
\E^A_M (x)
 \bd x^M \ ,\ \ \ \   \eq{5.15} $$
$$ \E^A_M =  C^A_{ \ B }(x) E^B_M \ , \ \ \ \ C_{AB} = \Tr ( T_A g T_B g\inv )
\ , \eq{5.16} $$
 we can represent (5.13) in the \sm form\foot {
Note that we  rescale the metric by the factor $-(k+\ha c_G)$ with respect to
the standard
definition (used in Sec.2). In particular, $G_{MN}$ is negative definite in the
 compact case.}
 $$
\G_{loc}(g)= S(x) = - {1 \over \pi \a' } \int d^2 z \ {\cal G}_{MN} (x) \del
x^M \bd x^N \ , \ \ \ \
\a' = {2\ov k + \ha c_G} \ , \eq{5.17} $$
 $$ G_{MN} \equiv  {\cal G}_{(MN )} = { G}_{0 MN }  -
\ha [K  (\M^{-1}- M^{-1}_{S})K^T]_{ab} E^a_M E^b_N   $$ $$ -
\ha (\M^{-1}- M^{-1}_{S})_{ab} \E^a_M \E^b_N -
 (K  \M^{-1}+ M^{-1}_{S})_{ab} E^a_{(M }\E^b_{N)} \ , \eq{5.18} $$
$$  B_{MN} \equiv  {\cal G}_{[MN ]} = B_{0 MN} -
 (K  \M^{-1} + M^{-1}_{S})_{ab} E^a_{[M }\E^b_{N] } \ . \eq{5.19} $$
   We have used that $K
M^{-1}_{S}K^T = M^{-1}_{S}$.
 Here ${\cal G}_{0MN }$  stands for  the  original  WZW  coupling,
$$ G_{0MN} = E^A_M \eta_{AB} E^B_N =  \E^A_M \eta_{AB} \E^B_N \ , \eq{5.20} $$
$$ 3\del_{[K} B_{0MN]} =  E^A_KE^B_ME^C_N f_{ABC} =\E^A_K\E^B_M\E^C_N f_{ABC}
\ .   $$
The  result  for the metric (5.18) is  equivalent to  the one  found  in \bs\
using the
`one-dimensional'
 ansatz. Our   direct  $2d$  approach   makes it possible  to derive also the
antisymmetric tensor
coupling  (5.19).\foot { An equivalent  expression was also   suggested  in
\bs\
using the analogy
with the result for the metric.}  It is useful to repeat the procedure of  the
elimination of  the gauge field  starting  directly  with the  original form of
the truncated
effective action (4.17),(4.18). Representing (4.18) as
 $$  \D I (g, A)= {1\over \pi }
 \int d^2 z \  \big[ (  - A\J  +
 \bar A J)   +    A  N \A
 +  \ha  b  \ ( AQA + \A Q^{-1} \A )   \big]  \ , \eq{5.21} $$
 $$   J_a =  \Tr (T_a g\inv\del g) \ , \  \ \ \J_a = \Tr (T_a \bd g g\inv ) \ ,
\ \ \ \  N_{ab}
\equiv M_{ab}  - b  \eta_{ab}\ , $$ and  solving for  $A,\A$   we obtain (after
omitting the  explicitly non-local terms where $Q$ or $Q^{-1}$ are  acting on
$N$)
$$   \D I (g) ={1\over 2 \pi }
 \int d^2 z \ \big[ J \V^{-1 } (N^T \J + b Q J)+ \J V^{-1 T} (N J + b
Q^{-1}\J)\big]   \ ,
 \eq{5.22} $$
$$ V\equiv N N^T - b^2  = MM^T- 2bM_S\ , \ \ \ \ \V\equiv N^T N - b^2= M^TM
-2bM_S \ ,\eq{5.23} $$
$$  \V^{-1} N^T = N^T V^{-1} \ , \ \ \ \M = M_S (MM^T)^{-1} V \ , \ \ \ \ K
V^{-1} K^T =
\V^{-1}  \ . $$
Ignoring  the non-local  terms  we can replace  $ Q^{-1}\J$ by $ \Tr (T_a \del
g g\inv ) $
and  $ QJ$ by $ \Tr (T_a g\inv\bd g ) $.   Comparing (4.17),(5.22) with (5.17)
and using
(5.14),(5.15) we find
$$ G_{MN}  = { G}_{0MN }  -
 b  (\V^{-1 })_{ab} E^a_M E^b_N -
  b (V^{-1 })_{ab} \E^a_M \E^b_N -
 2 (  \V^{-1 } N^T )_{ab} E^a_{(M} \E^b_{N)}\ , \eq{5.24} $$
$$  B_{MN}  = B_{0MN} -
 2 (  \V^{-1 } N^T )_{ab} E^a_{[M} \E^b_{N]}\ . \eq{5.25} $$
These expressions  \bs\  can be   transformed into the form
(5.18),(5.19)  with the help of (5.23).

Given the effective action (4.17),(4.18) it is  possible also to derive  the
expression for the dilaton coupling.
 As was suggested in \wit\ and discussed in detail in \tsw\
the dilaton  contribution can be found from the regularised determinant which
appears after one
integrates over the gauge field.  It should be emphasized that we are not
actually treating the
arguments $A,\A$ of the  effective action as quantum fields. The correct point
of view  is that the
dilaton coupling term  should be already contained in the quantum effective
action (1.1) computed on
a curved  $2d$ background.  The reason why   the exact expression for the
dilaton can be found
from the determinant of the $A,\A$ - bilinear form   is related to   the fact
that the dilaton term
can be interpreted be  as an anomaly-type  (`semiclassical')  contribution.
If
$$ Z= \int [dA_\a] \exp {( - \2p \int d^2z \sqrt \g \  F_{\a\b} A_\a A_\b  )} \
, \eq{5.26 } $$
where $F_{\a\b}(z) $ is a given matrix function  and $\a, \b $ stand for both
internal and
two-dimensional indices, then
 $$ Z = \exp [ - \ha ({\rm Tr }  {\ \rm ln \  }  F  )_{reg} ]  $$ $$ =
\exp [ -{1\over { 16 \pi  }} \int
d^2 z \sqrt {\g} \ [
 c_0 \Lambda^2   {\ \rm ln \  det \ }  F   +  c_1
(\del_m  {\ \rm ln \  det \ }  F )^2  \  +  \  c_2  R  {\ \rm ln \ det \ }  F
\ ] \ \ . \
\eq{5.27 } $$  Here $\Lambda $ is an UV cut-off, $R$ is the  curvature  of  the
$2d$
metric  $\g_{mn}$ and
$c_i$ are finite coefficients.
 To   preserve   the conformal invariance of the theory one should define $Z$
in such a way
that   $c_1=0$ and $c_2= -1$ \bush\mplt\shts.\foot {
The $c_1$-term would have given a correction to the \sm metric. In the
semiclassical
approximation one could argue that such term should be dropped since  the
corresponding correction
to the metric would contain an extra power of $\a'$. }
g The quadratically divergent  term can be
interpreted  as a contribution to   the   local measure     while the
coefficient of $R$ is the
dilaton coupling of the corresponding   sigma model, i.e.
 $$ Z= \prod_z  {\det F}^{-1/2} \ \exp \  [  \ - {1\over {
4 \pi  }} \int d^2 z\  \sqrt {\g}  \ R \ \p \ ] \ \ ,  \ \ \eq{5.28} $$
$$ \p    =  - \fourth    {\ \rm ln \ det \ }   F  \ .  \eq{5.29 } $$
Taking into account the Jacobian  of the transformation (5.6) (equal to  $\det
P$) we get
from (5.7) $$  \det F =  \det \M \  \det M_S   \ (\det P)^{-2} \ . \eq{5.30 }
$$
Using that
$$ \M =  M_S (MM^T)^{-1} V\ ,    \ \ \ \
\det M_S =  \det M \det P  \ , \eq{5.31  } $$
we  find
$$ \p    =  - \ha \ {\rm ln \ det } \  M  +  \fourth   \ {\rm ln \ det } \  M_S
  - \fourth \
{\rm ln\  det } \   \M  \
      \eq{5.32  } $$
$$  =
 - \fourth  {\ \rm ln \ det \ }  V   \ . \eq{5.33 } $$
A similar representation for the exact dilaton \dvv\ in the $SL(2,R)/U(1)$
model
was  given   in \tsw, while the  expression  equivalent to (5.33)  has also
appeared in \bs.

The \sm with the metric (5.18), antisymmetric tensor (5.19) and dilaton (5.32)
should  be   conformal
invariant to all orders in  the \sm loop expansion, i.e. should represent a
large
class of exact
solutions of string equations.  Depending on $b$, the  fields $G_{MN}, B_{MN}$
and $\p$ are
non-trivial functions of  parameter $k$ or  $\a'$ (see (5.17))
$$  k +\ha c_G = {2\ov \a'}   \ \ , \ \ \ \ \ \ b= -  \fourth (c_G - c_H) \a'
\ ,
 \eq{5.34} $$
i.e. the semiclassical limit corresponds to $b \ra 0$.

As we have found in Sec.4.2,  the  local part  of  the bosonic term in  the
effective action in the
gauged supersymmetric WZW theory  is  equal to the classical bosonic gauged WZW
action (4.37) with
unshifted $k$ and no `quantum' $b$-term. Thus  $\a' = {2\ov k}$ and     the
corresponding  exact \sm couplings are given   by  the `semiclassical' bosonic
expressions
(5.18),(5.19),(5.32) with $b=0$, i.e.
 $$ G_{MN}^{(s)}  = { G}_{0 MN }  -
 2 ( M^{-1})_{ab} E^a_{(M }\E^b_{N)} \ ,
  \eq{5.35} $$
$$ B_{MN}^{(s)}  = B_{0 MN} -
 2( M^{-1})_{ab} E^a_{[M }\E^b_{N] } \ , \ \ \ \ \ \  \p^{(s)}   =  - \ha  {\
\rm ln \ det \ } M \ .
\eq{5.36 } $$   For example, in the case when $G/H$ is K\"ahler,  (5.35)  and
(5.36)
give the couplings of
the \sm  corresponding to $N=2$  Kazama-Suzuki superconformal theories (both
compact \ks\ and
non-compact \bb\bn\bsf\bsfet).  As was already anticipated in \ks\ the \sm
metric is  different from
the  invariant  metric on $G/H$. It is  now clear that this can be attributed
to the presence of a
non-trivial dilaton coupling.
\subsec{Explicit  form of  the  sigma model  metric   }
 To be able to study the
properties of (5.18),(5.19),(5.32), and, in particular,
to establish the  correspondence with the results found in the operator
formalism in Sec.2 (e.g. to
prove that  the  dilaton can be expressed in terms of the ratio of the
determinants of
the `deformed' and invariant metrics on $G/H$ (2.22))  it is important to
transform the metric (5.18)
into a more explicit form.
 This, in fact, can be done in general (without  specifying $G$ and $H$).
  First, let us note that $\E^A_M$ can be
expressed in terms of $E^A_M$  with the help of the  matrix $C_{AB}$ (5.16),
$$ \E_{aM} = C_{ab} E^b_M + C_{ai} E^i_M \ , \ \ \  \E^M_a = C_{ab} E^{bM} +
C_{ai} E^{iM} \ ,
\eq{5.37}  $$
$$ E^A_M E^M_B = \delta^A_B \ , \ \ \ E^A_M E^N_A = \delta^N_M \ , \ \ \
 \E^A_M \E^M_B = \delta^A_B\ , \ \ \ \E^A_M \E^N_A = \delta^N_M  \ ,  \ \ \
\E^A_M E^M_B= C^A_{\
B} \ . $$ The  indices are raised and lowered with  $\eta_{AB}$ and $G_{0MN}=
E^A_ME_{AN}=
E^a_ME_{aN}+ E^i_ME_{iN}$.
The matrix $C_{AB}$ satisfies the orthogonality relation $ C^T \eta C = \eta $,
i.e.,   in
particular,
   $$   C_{ad} C_{b}^{\ d } + C_{ai} C_{b}^{\ i}  = \eta_{ab} \ .  \eq{5.38} $$
Using  these relations  we can put (5.18),(5.24) into the form
$$ G_{MN}  =  h_{AB} E^A_M E^B_N   =
  h_{ij} E^i_M E^j_N +  h_{ab} E^a_M E^b_N
  + 2h_{ai} E^a_{(M }E^i_{N)} \ , \eq{5.39} $$
where
$$
 h_{ij }= \eta_{ij} +  f_{ab}  C^a_{\ i} C^b_{\ j}\ , \ \ \
f_{ab} = -   b  (V^{-1 })_{ab} =  -\ha (\M^{-1}- M^{-1}_{S})_{ab} \ ,
\eq{5.40} $$
$$  h_{ab }= \eta_{ab} - b \V^{-1}_{ab} - b V^{-1}_{cd}C^c_{\ a} C^d_{\ b}  -
(  \V^{-1 }
N^T)_{ad}  C^d_{\ b} -   (  \V^{-1 }
N^T)_{bd}  C^d_{\ a} \ ,
 $$
$$  h_{ai }=  - b V^{-1}_{bd} C^b_{\ a} C^d_{\ i}   -  (  \V^{-1 } N^T  )_{ad}
C^d_{\ i} \ . $$
 The key observation is  that
this metric is {\it degenerate}, having  $D_H$ null vectors  $$ G_{MN} Y^N = 0
\ , \ \ \  \  Y^N
=E^N_A Y^A \ , \ \  Y^A = \{ y^a \ , \ \ y^a M^{-1}_{ab} C^{bi} \} \ ,
\eq{5.41} $$
where $y^a$ are free parameters (this is true,  of course,  for an arbitrary
value of $b$, i.e. also
for the  `semiclassical' metric  (5.35)).   In view of  (5.37) and  the
relation $C_{ab} = M_{ab} +
\eta_{ab}$ ,
 $$  Y^N = - y^a M^{-1 b}_{a\ }
( E^N_b - \E^N_b)  , \eq{5.42} $$
 i.e. the equivalent set of null vectors is    represented by
$$    Z^N_a =  E^N_a - \E^N_a = - M_{ab} E^{Nb}- C_{ai} E^{Ni} \ . \eq{5.43} $$
 These vectors  are recognised as being the generators of the vector
subgroup $H$ of  the $G\times G$ symmetry of the WZW  action  which was gauged
in (4.1).
 They  are
Killing vectors of the group metric $G_{0MN}$ and are also  the Killing vectors
 of $G_{MN}$ (see in this
connection \jjmo).
Note that  it is the `$Z$-component' of the current ${\tilde {\cal J}}_a$
 (5.10)  which  is
coupled to the
`longitudinal' part $\bar B$  of the gauge potential in (5.11).

It is  possible   to check that $G_{MN} Z^N_a = 0$  directly  using (5.43), the
representation
(5.24) for the metric and  (5.37),(5.38).
As a simple illustration,   let us consider a special
case of symmetric $C_{ab}$ ($C$ is symmetric, for example, when ${\rm dim} \  H
 =1$). Then
$$ M_S = M\ ,  \  \ \ K = P= 1  \ ,  \ \ \M = M - 2 b  \ , \ \ \ V=\V= M (M-2b)
 $$
 so that the background fields (5.18),(5.19),(5.32)
take the form $$ G_{MN}  =
 { G}_{0MN }  - 2 (M^{-1})_{ab} E^a_{(M} \E^b_{N)} -
 b  [M(M-2b)]^{-1}_{ab} (E^a_M  + \E^a_M )(E^b_N  + \E^b_N)  \  ,
  \eq{5.44 } $$
$$  B_{MN}  = B_{0MN} -
 2 \{[M(M-2b)]^{-1} (M-b)\}_{ab} E^a_{[M} \E^b_{N]}\ , \eq{5.45 } $$
 $$ \p    =  - \fourth    {\ \rm ln \  det \ } [  M   (M - 2b ) ]
\ .    \eq{5.46 }  $$
In (5.44) we have separated the `semiclassical' part from the higher-order
correction.
Multiplying  (5.44) by $E^N_a - \E^N_a$ and using that
$$ (E_{aN } + \E_{aN}) (E^N_b - \E^N_b) =  C_{ab} - C_{ba}  \  \eq{5.47} $$
vanishes  by  assumption,   one  concludes that both the `semiclassical' and
`quantum' parts of
(5.44) give zero contributions  to the product.

 To get a non-degenerate metric we should restrict  $G_{MN}$  to the subspace
orthogonal
to  the null vectors $Z^N_a$. In general, given a set of null vectors and
another  non-degenerate
`canonical' metric (which we shall choose to be equal to the invariant metric
$G_{0MN}=E^A_M E_{AN}$
on $G$)  one  can define
 the projection operator on the subspace orthogonal (with respect to $G_{0MN}$)
to $Z^M_a$
$$ \Pi^N_M
\equiv  \delta^N_M -  Z^N_a (ZZ)^{-1ab} Z_{Mb} \ , \ \ \ \ (Z
Z)_{ab} = G_{0MN} Z^M_a Z^N_{b}\ , \ \ \ \  \Pi^2 = \Pi \ . \eq{5.48} $$
 One  can  change  the original basis $E^M_A = (E^M_i , E^M_a)$  for the  new
one
$ (H^M_i, Z^M_a)$  with $H^M_i$ being  orthogonal to $Z^M_a$
$$  G_{0MN} H^M_i Z^N_a= 0 \ , \ \ \  \ \Pi^N_M H^M_i = H^N_i \ . \eq{5.49} $$
Then  the  degenerate metric (5.39)  takes the form ($H^i_M \equiv H^N_j
\eta^{ij} G_{0MN}$)
$$  G_{MN} = \Pi_M^K {\hat G}_{KL} \Pi^L_N
\ , \ \ \ \ {\hat G}_{KL} =
  g_{ij} H^i_M H^j_N +  g_{ab} Z^a_M Z^b_N
  + 2g_{ai} Z^a_{(M }H^j_{N)} \ , \eq{5.50} $$
i.e.
$$ G_{MN}  =  g_{ij} H^i_M  H^j_N \ . \eq{5.51} $$
The choice of $H^M_i$ is not unique.  We  can  take
 $$H^M_i = \Lambda_i^j {\tH}^M_j \ , \ \ \ \ { \tH}^M_i = \Pi^M_N E^N_i \  , \
\ \ \
 (HH)_{ij}=  G_{0MN} H^M_i H^N_j \ , \eq{5.52} $$
where $\Lambda_i^j $ can be fixed, for example, by  the condition of
orthonormality of $H^M_i$, i.e. $(HH)_{ij}= \eta_{ij}$.   The inverse  to
$G_{MN}$  is given by
$$ G^{-1MN}G_{NK} = \Pi^M_K \ , \ \ \  \ G^{-1MN}=  g^{(-1) ij} H_i^M  H_j^N \
, \ \ \
    \Pi^N_M =     H^N_i (HH)^{-1ij} H_{Mj}         \ , \eq{5.53} $$
$$    g^{(-1)ik} (HH)_k^l g_{jl} = (HH)^{-1i}_{\ k }\ . \eq{5.54} $$
In our   case of
(5.43), $ (ZZ)_{ab}  = -2 (M_{S})_{ab}$, i.e.,
$$ \Pi^{N}_M  =  \d^N_M + \ha (E^N_a - \E^N_a) M_{S}^{-1 ab} (E_{Ma} - \E_{Ma})
\ .  $$
To express the metric in terms of $\Pi^{N}_M$ the form (5.18) of $G_{MN}$ is
most useful:
as  is obvious from (5.18),
$$ G_{MN} =  \Pi_{MN} - \ha (\E_{Ma} + E_{Mc}K^c_a) \M^{-1 ab} (\E_{Nb}  +
K_b^c E_{Nc}) \ . $$
Since
$ \Pi^{N}_M E^M_a =  \ha ( \E^{Nb} + E^N_cK^{cb}) (M_S^{-1} M)_{ba} \ ,   $
we find (5.50) with
$$
{\hat G}_{MN} =   G_{0MN} -  2 E_{M}^{a} \M^{-1}_{ ab} E_{N}^b   =
 E_M^i\eta_{ij} E_{N}^{j}  +  E_{M}^{a}( \eta_{ab} - 2  \M^{-1}_{ ab})  E_{N}^b
\ .\eq{5.55} $$
A  simple  choice of (non-orthonormal) $H^M_i$ is
(we shall use bars to denote objects  corresponding  to this basis)
$$  \bH^i_M = E^i_M - M^{-1}_{ab} C^{bi} E^a_M = p^i_j E^j_M +
M^{-1}_{ab} C^{bi} M^{-1a}_{\ c} Z^c_M\ , \ \ \ \    \bH^i_M Z^M_a = 0 \ ,
\eq{5.56} $$
$$(\bH\bH)_{ij} = \bH_{iM} \bH_j^M = p_{ij} \ , \ \ \ \ p_{ij} \equiv
\eta_{ij} +
(MM^T)^{-1}_{ab}  C^a_{\ i} C^b_{\ j}\ .
 \eq{5.57} $$
Expressing $E^i_M, E^a_M $ in terms of $\bH^i_M, Z^a_M$ is effectively
equivalent to  dropping out
terms with $E^a_M$ in (5.39), i.e.,  we find
 $$G_{MN}    =
  {\bar g}_{ij} \bH^i_M \bH^j_N\ , \ \ \ \
{\bar g}_{ij} = h_{ij} = \eta_{ij} -   b  V^{-1 }_{ab} C^a_{\ i} C^b_{\ j}\ . \
 \eq{5.58} $$
Note that the metric ${\bar g}_{ij}$ becomes  trivial in the `semiclassical'
limit $b=0$.
If instead we use the basis $\tH^i_M$ defined in (5.52)
$$ \tH^i_M = \Pi^M_N E^N_i = E^i_M +  p^{-1}_{ij}  C^{a j} (MM^T)^{-1 }_{ab}
Z^b_M\ , \eq{5.59} $$
then
$$ E^a_M= - M^{-1ab} C_{b i} \tH^i_M  - ( \eta^{ab} -  M^{-1ac}  M^{-1bd}
C_{c i} p^{-1ij}  C_{d j}) Z_{bM} \ , $$
and thus from (5.55)
$$G_{MN} =
  {\tilde g}_{ij} \tH^i_M \tH^j_N\ ,
\ \ \ \ {\tilde g}_{ij}= \eta_{ij}  + [ M^{-T} (1- 2\M^{-1})M^{-1}]_{ab}
C^a_{\ i} C^b_{\ j}\ . \
\eq{5.60} $$
The representations (5.58) and (5.60) are related by the transformation
$$ \tH_M^i = (p^{-1})^i_j \bH_M^j \ , \ \ \ \ {\bar g}_{ij}= {\tilde g}_{kl}
(p^{-1})^k_i
(p^{-1})^l_j\ ,  $$
$$  (\tH \tH)_{ij} = p^{-1}_{ij} = \eta_{ij} +  \ha (M^{-1}_S)_{ab}  C^a_{\ i}
C^b_{\ j}\ .
\eq{5.61}
$$
The  $D\times D$ matrices  of generic form  $h_{ij }= \eta_{ij} +  f_{ab}
C^a_{\
i} C^b_{\ j}$   have the following  multiplication law:
$$( \eta_{ik} +  f_{1ab}  C^a_{\ i} C^b_{\ k})\eta^{kl} ( \eta_{lj} +  f_{2ab}
C^a_{\ l} C^b_{\ j})
= ( \eta_{ij} +  f_{3ab}  C^a_{\ i} C^b_{\ j}) \ , \ \ \ $$
$$ f_{3 }= f_{1} + f_{2} + f_{1} ( 1- CC^T) f_2\ , \eq{5.62} $$
where we have used (5.38), i.e. that $C_{a}^{\ i }C_{bi}= (1-CC^T)_{ab}$. The
inverse of such a
matrix is given by $$ h^{-1}_{ij} = \eta_{ij} + f^{(-1)}_{ab}  C^a_{\ i} C^b_{\
j}\ , \ \ \ \
f^{(-1)}=- [f^{-1}  +  ( 1- CC^T) ]^{-1}. \eq{5.63} $$
  The determinant of  $h_{ij}$ can be computed by taking the variation with
respect to
 $f_{ab}$
 $$ \d {\  \rm ln \ det}\  h = \Tr ( h^{-1} \d h) =  h^{-1ij} C_{ai} C_{bj} \
\d f^{ab}
= \{( 1- CC^T)[ 1 + f^{(-1)} ( 1- CC^T)]\}^{ab}\ \d f^{ab}\ , $$
$$  \det h =  \det [1+ f(1-CC^T)]  \ \ . \eq{5.64} $$
Applying (5.64) to $p_{ij}$, ${\bar g}_{ij}$ and ${\tilde g}_{ij}$  in
(5.57),(5.58) and (5.60) we
get
$$
\det {\bar g}_{ij} = \det [ (1+b) V^{-1} M^2] \ , \ \ \ \det p_{ij}=  \det [-2
M^{-2}M_S]  \
, \eq{5.65} $$
 $$  \det {\tilde g}_{ij}= \det [4(1+b) V^{-1}  M_S^2 M^{-2} ] =  \det {\bar
g}_{ij}\
(\det p_{ij})^2 \  . \eq{5.66} $$
The inverse of $G_{MN}$ defined according to (5.53),(5.54) is found to be
$$  G^{-1MN}=  {\bar g}^{(-1)ij} \bH_i^M  \bH_j^N \ ,
\ \ \ \ {\bar g}^{(-1)}_{ij}  \equiv (p{\bar g}p)^{-1}_{ij} \ , \eq{5.67} $$
$$ {\bar g}^{(-1)}_{ij}=   \eta_{ij} +  \ha [  M_S^{-1}- {1 \ov 2
(b+1)}M_S^{-1}MM^T
M_S^{-1}]_{ab}  C^a_{\ i} C^b_{\ j} \ . \eq{5.68} $$
The metric (5.67) can be represented in the following form
$$  G^{-1MN}=  \Pi^M_K {\hat G}^{-1KL} \Pi_L^N \ ,  $$ $$ \ \ \ \
{\hat G}^{-1KL}=   E^M_AE^{AN} -  \g   E^M_aE^{aN}
= E^M_iE^{iN}  -   (\g -1)   E^M_aE^{aN}\ ,  \eq{5.69}  $$
$$
 \g = (b+1)^{-1} = {k+\ha c_G\over  k+\ha c_H } \ . \eq{5.70} $$
This is readily checked  using (5.43),(5.56),  i.e.
$$E^i_M = p^{-1i}_{\ j } \bH^j_M + O(Z)\ , \  \ \   E^a_M =  - M^{-1ab} C_{b i}
p^{-1i}_j
\bH^j_M + O(Z)\ ,  $$
$$ E^M_iE^{iN}  -   (\g -1)   E^M_aE^{aN} = (p^{-1}g'p^{-1})_{ij} \bH^M_i
\bH^{iN} \ , $$
$$ g'_{ij} = \eta_{ij} - (\g -1) (MM^T)^{-1}_{ab} C^a_{\ i} C^b_{\ j} \  , \ \
\ \ \ \ g' = {\bar
g}^{-1} \ ,  \eq{5.71} $$
 where  we have noted that the  matrix  $g'$  is nothing but the inverse of
$\bar g$ in (5.58).

To  obtain a similar   representation for the antisymmetric tensor
coupling  in (5.25)
one is to express $ \E^a_M$ in terms of $E^i_M, E^a_M$ (or $\bH^i_M, Z^a_M$).
  We  get
$$  B_{MN}  = B_{0MN} -
 2 (\V^{-1 } N^T  C)_{ab} E^a_{[M} E^b_{N]}
-2 (\V^{-1 } N^T)_{ab}C^b_{\ i}  E^a_{[M} E^i_{N]}
\ . \eq{5.72} $$
 It should be possible to find a more explicit  representation  for the
field   strength of $B_{MN}$  using the expression for $\del_{[K} B_{0MN]}$.
The gauge invariance of the action (before gauge fixing)
implies  that the Lie derivative of $H_{MNK}$ along $Z^M_a$ should vanish and
that $B_{MN}
Z^M_a=0$, i.e.  like $G_{MN}$ the antisymmetric tensor  can be  represented  in
terms of $\H^i_M$.

Since the \sm  $ \int d^2z  G_{MN} (x) \del  x^M \bd x^N  + ... $ has a gauge
invariance (generated
by $Z^M_a$) the   final step  is to  fix a gauge, e.g.  to  restrict
coordinates  $x^M$ on $G$ to
coordinates  $x^\m$ on $G/H$.  Let  $$ R^a (x^M) =0 \ , \ \ \ \  R^a_M \delta
x^M  = 0 \ , \ \ \
R^a_M \equiv  \del_M R^a  \  \eq{5.73}       $$
 be  a gauge condition (the corresponding ghost determinant  which should be
included in the
measure is  $\det R^a_M Z^M_b $).  One  may  either  add a gauge term into the
\sm action
(which will then  depend on  all $x^M$ coordinates) or explicitly solve the
gauge condition
 expressing  $x^M = x^M (x^\m )$ in terms of $D$ coordinates $x^\m$ on
$G/H$.\foot
{If one uses  the formulation in terms of all  $D_G$ coordinates $x^M$
one should  also impose as usual the gauge invariance    (BRST invariance)
condition  on the
observables. Adding a gauge-fixing term in the action one  obtains the
following  non-degenerate
metric on  $G$: $\  {\bar
 G}_{MN} = G_{MN} + q_{ab} R^a_M R^b_N$, where $q_{ab}$   can be chosen  on the
basis of
convenience.
 The  determinant of the degenerate metric  $G_{MN} $ is then  formally defined
as
follows $$  (\det { G_{MN}})^{-1/2} =  (\det
 {{\bar G}_{MN}})^{-1/2} \ \det (R^a_M Z^M_b)  (\det q_{ab})^{1/2}\
. $$ }
In the latter case we  will get a \sm on the $D$-dimensional space  with
 $H^i_M$   replaced by  the $D\times D $
matrix   $H^i_\m \equiv H^i_M \del_\m x^M $ (i.e.  a   vielbein),  $G_{MN}$ by
$G_{\m\n}$,  etc.
The  expressions for the  \sm metric  and antisymmetric tensor    then are
$$ G_{\m\n}    = G_{MN}\del_\m x^M \del_\n x^N \  ,
\ \ \ \   G_{\m\n}    = G_{MN}\del_\m x^M \del_\n x^N \
   ,\eq{5.74}
$$
where $G_{MN}$ and  $B_{MN}$  are given by (5.58),(5.56) and (5.72).
\newsec {Equivalence of   Results  of  Operator  and  Field--Theoretical
Approaches   }
Let us now compare the  expressions  for the background metric and the dilaton
corresponding  to a gauged WZW model  obtained   in the field-theoretical
approach
(5.58),(5.73),(5.33)
$$  G_{\m\n}    =
  {\bar g}_{ij} \bH^i_\m \bH^j_\n \ , \ \ \ \
{\bar g}_{ij} = \eta_{ij} -   b  V^{-1}_{ab} C^a_{\ i} C^b_{\ j}\ . \  \eq{6.1}
$$
$$ \p = \p_0
 - \fourth  {\ \rm ln \ det \ }  V   \ , \ \ \ \ \
    V = MM^T - b(M+M^T) \ , \ \ \ \  \eq{6.2 } $$
$$ M_{ab } = \Tr (T_a g T_b g\inv  - T_a T_b ) \ ,\ \ \ \ C_{ai} =  \Tr (T_a g
T_i g\inv )\  , $$
 with the
results   (2.12),(2.25),(2.31) which were  found  in Sec.2  by identifying the
operator  $L_0$
of conformal $G/H$ theory with a
Klein-Gordon operator in a background.  The restriction  of $L_0$ to
$H$-invariant
 states
 implies  imposing  the  condition $Z^M_a \del_M =0$ (where $Z^M_a$ is  given
by  (5.43)). Then   the
metric (2.12) is the `projected' one (2.25)  and therefore  is {\it equal } (up
to the overall factor
$-(k+\ha c_G)$ which we separated from the \sm  metric in (5.17))  to the
inverse of the metric
(5.73)  (see (5.69),(5.70)). We have thus demonstrated that  both the operator
and the
field-theoretical  approaches  lead to the same expression for the target space
metric. While the
operator approach gave the inverse of the metric  in a simply looking  but
rather abstract
form,  following the direct \sm  approach   provided  us with   more explicit
representation  for
the metric itself and clarified its general structure.

The metric (6.1) can be considered as  a
 `deformation' of the `round' metric on $G/H$.
The latter  corresponds to the \sm which is found by integrating out the gauge
field  (taking
values in the algebra of $H$) in the  action
 invariant under the gauge transformations $g'=gu$ generated by $E^a_M$
 $$ I = \int d^2 z\   \Tr (
g\inv \del_m g +  A_m )^2  \ . \eq{6.4}
$$
This action (and hence the resulting \sm metric) has also global $G$-invariance
which  is {\it absent} in the gauged WZW action (4.1)  (unless $H$ is an
invariant subgroup of
non-simple $G$)  being broken by the $g^{-1}Ag\A$-term). As a result,  in
contrast to the metric in
(5.74), the  \sm metric corresponding to (6.4)  has global $G$ invariance.
 Before gauge fixing we get a degenerate metric
$G^{(0)}_{MN}$ on the  full $G$
 (with null vectors
$E^M_a$). Solving  a gauge condition  (5.73)
 and expressing $x^M$ in terms of $x^\m$ we get
the  metric $G^{(0)}_{\m\n}$ on  $D$-dimensional space $G/H$,
$$ G^{(0)}_{MN} =\eta_{ij}E^i_M E^j_N \ , \ \ \ \  G^{(0)}_{\m\n}=
\eta_{ij}E^i_\m E^j_\n  \ , \ \ \ E^i_\m = E^i_M \del_\m x^M
 \ . \eq{6.5} $$
Note that since  according to (5.56) $ \bH^i_M = E^i_M + O(E^a_M)$,   choosing
the gauge condition
 in (5.73),(5.74) such that $ \del_M R^a = E^a_M $
 it is easy to  see  that
$$ \det G_{\m\n} =   \det G^{(0)}_{\m\n} \ \det {\bar g}_{ij} \ (\det M)^{-2}\
,  \eq{6.6} $$
or  (cf. (5.35))
$$ \det G^{(s)}_{\m\n}  = \det G^{(0)}_{\m\n}
  \  (\det M)^{-2}  \ ,  \ \ \ \
 G^{(s)}_{\m\n}= {\eta}_{ij} \bH^i_\m \bH^j_\n \ ,\eq{6.7} $$
where $ (\det M)^2 $  is the  square of the  corresponding ghost determinant
($E^a_M Z^M_b = -
M^a_b $ , see (5.43)).

A similar equivalence is  established   between  the  results of the two
approaches for the dilaton
(2.31) and (5.33). As we have shown, the operator approach implies that the
dilaton is given   by the
logarithm of the ratio (2.31) of the determinant of the  metric and the
determinant of an invariant
metric on the coset space.   Using  (6.1),(6.2) and the expression (5.65) for
the determinant of
${\bar g}_{ij}$  and  we get
$$ \sqrt { \det G_{\m\n} }  \ \e{-2 \p} =
 \sqrt {  \det G^{(s)}_{\m\n} } \ \sqrt { \det {\bar g}_{ij} }
 \ \e{-2 \p}  \eq{6.8} $$ $$
= \  a \  \sqrt {
 \det G^{(s)}_{\m\n} } \ \det M  =\  a  \ \sqrt {  \det G^{(0)}_{\m\n} } \  ,
\eq{6.9} $$ $$  a =  (\sqrt {1+b} )^{D_H}\e{-2 \p_0} \ , $$
i.e. as in  (2.22)  the `measure factor' (6.8) is  nothing but the
canonical  measure on the  $G/H$.
 Since $G^{(0)}_{\m\n}$ is $b$-independent we  have
thus proved  in general that
$ \sqrt G \ {\rm e}^{-2 \p}$
 is essentially $k$-independent, in agreement with
 \kir\bsfet\bs.

  It would be interesting to study the
 general properties  of the metric (6.1)
, relating them to the properties of the basic matrix
$C_{AB}$ (i.e. of  $M_{ab}, C_{ai}$, etc)   and the value of $b$ (for example,
in
the semiclassical limit
  the singularities of (5.35),(5.36)  correspond to the fixed points of the
transformation $g\ra hgh\inv$ where $M\ra 0$ \ginq; see also \bsft\bsfet).

 The \sm couplings in the supersymmetric case (5.35),(5.36) does not depend on
$\a'$. That means they  solve the  \sm conformal invariance conditions
 at each order of the loop expansion. Namely, once the one-loop  conditions are
satisfied,   all
higher loop corrections to the  $\beta$-functions
 vanish  (up to a field redefinition ambiguity) on
the corresponding   background.\foot { This  was checked explicitly  up to
5-loop order in the
supersymmetric $SL(2,R)/U(1)$ model
 \Jack. While the  two -  and three - loop   terms in the $\b$-function of the
$N=1$ supersymmetric
\sm  are known to  vanish  (in the minimal subtraction scheme) \alv, the
four-loop
term does not vanish  in general \gr.  However, there exists such a
renormalisation scheme in which it
vanishes for the  `one-loop' $D=2$ background of \efr\wit. }  This is
interesting, given that these
models  have, in general, only $N=1$ (and not $N=2$ or $N=4$)  supersymmetry.
There exists  another
class   of $N=1$ supersymmetric  sigma models  which have  the same  property
\tttt.\foot { The
corresponding     $2+D$-dimensional  target space metric  has  null Killing
vector   and the
`transverse' $D$-dimensional part of the space  is represented by  the
homogeneous  $G/H$
K\"ahler-Einstein manifolds (the  $N=2$  sigma models  representing   the
`transverse' space have
only one-loop  $\b$-function being non-zero \per). }

\bigskip\bigskip\bigskip
I am grateful to  I. Bars,  A.M. Semikhatov, K. Sfetsos  and A. Turbiner for
discussions and
 critical remarks.  This work was  partially supported by a grant from SERC.

\listrefs
\end

\vfill\eject
\listrefs
\end